\documentclass[11pt,a4paper]{scrartcl}
\setkomafont{caption}{\sffamily}
\usepackage[ansinew]{inputenc}
\usepackage[british]{babel}
\usepackage[scale=0.72]{geometry}
\usepackage{fancyhdr}
\usepackage{graphicx}
\usepackage{hepunits}
\usepackage{hyperref}
\usepackage{eurosym}
\usepackage{pof3_defs}

\usepackage{SIunits}
\usepackage{subfig}

\graphicspath{{./}}

\begin{document}
\setcapindent{0mm}

\sffamily

\title{Enabling Technologies for Silicon Microstrip Tracking Detectors at the HL-LHC}
 
\author{\huge The PETTL Collaboration\\ \\
L. Feld, W. Karpinski, K. Klein, M. Preuten,\\
J. Sammet, M. Wlochal\\
{\it RWTH Aachen University, 1. Physikalisches Institut B}\\ \\
H. Lacker, D. Sperlich, L. Rehnisch, M. Daniels\\
{\it Humboldt University Berlin}\\ \\
I. Bloch, G. Eckerlin, D. Eckstein, T. Eichhorn, \\
I.-M. Gregor, A. Mussgiller, L. Poley, M. Stanitzki\\
{\it DESY} \\ \\
C. A. Betancourt, M. Hauser, K. Jakobs, S. Kuehn, K. Mahboubi, U. Parzefall \\
{\it Albert-Ludwigs-University Freiburg, Institute of Physics}\\ \\
J. Erfle, E. Garutti, A. Junkes, T. Poehlsen,\\
P. Schleper, S. Schuwalow, G. Steinbrück\\
{\it Hamburg University}\\ \\
C. Barth, F. Bögelspacher, W. de Boer, A. Dierlamm, R. Eber, M. Guthoff, \\
F. Hartmann, U. Husemann, Th. Müller, A. Nürnberg\\
{\it Karlsruhe Institute of Technology, IEKP} 
 }

\date{\today}

\maketitle

\newpage

\noindent {\Large Abstract}\\
While the tracking detectors of the ATLAS and CMS experiments have shown excellent performance in Run 1 of LHC data taking, and are expected to continue to do so during LHC operation at design luminosity, both experiments will have to exchange their tracking systems when the LHC is upgraded to the high-luminosity LHC (HL-LHC) around the year 2024. The new tracking systems need to operate in an environment in which both the hit densities and the radiation damage will be about an order of magnitude higher than today. In addition, the new trackers need to contribute to the first level trigger in order to maintain a high data-taking efficiency for the interesting processes. Novel detector technologies have to be developed to meet these very challenging goals. The German groups active in the upgrades of the ATLAS and CMS tracking systems have formed a collaborative \textquotedblleft Project on Enabling Technologies for Silicon Microstrip Tracking Detectors at the HL-LHC\textquotedblright (PETTL), which was supported by the Helmholtz Alliance \textquotedblleft Physics at the Terascale\textquotedblright during the years 2013 and 2014. The aim of the project was to share experience and to work together on key areas of mutual interest during the R\&D phase of these upgrades. The project concentrated on five areas, namely exchange of experience, radiation hardness of silicon sensors, low mass system design, automated precision assembly procedures, and irradiations. This report summarizes the main achievements.\\ \\

\noindent {\Large Editors}\\
Alexander Dierlamm, Lutz Feld, Alexandra Junkes, Katja Klein, Susanne Kühn, Andreas Mussgiller, Ulrich Parzefall, Marius Preuten.\\ \\

\noindent {\Large Acknowledgements} \\
We thank the Helmholtz Alliance  \textquotedblleft Physics at the Terascale\textquotedblright for funding the PETTL-Project. This project brought ATLAS and CMS Tracker experts together, and resulted in the work presented in this document.\\
Of course, the PETTL-Project was embedded in the German ATLAS and CMS projects, which received funding from the BMBF and HGF. \\
The research leading to these results has received
funding from the European Commission under the FP7 Research
Infrastructures project AIDA, Grant agreement no. 262025.\\
The sensors used for the measurement shown in Figure 9 were produced within the RD50 project.\\
One author is co-financed by the European Social Fund and by the
Ministry Of Science, Research and the Arts Baden-Wuerttemberg.

\newpage 

%%%%%%%%%%%%%%%%%%%%%%%%%%%%%%%%%%%%%%%%%%%%%%%%%%%%%%%%%%%%%%%%%%%%%%%%%%%%
\tableofcontents
%%%%%%%%%%%%%%%%%%%%%%%%%%%%%%%%%%%%%%%%%%%%%%%%%%%%%%%%%%%%%%%%%%%%%%%%%%%%

\section{Introduction}
\label{sec:intro}

While the tracking detectors of the ATLAS and CMS experiments have shown excellent performance in Run 1 of LHC data taking, and are expected to continue to do so during LHC operation at design luminosity, both experiments will have to exchange their tracking systems when the LHC is upgraded to the high-luminosity LHC (HL-LHC) around the year 2024. The new tracking systems need to operate in an environment in which both the hit densities and the radiation damage will be about an order of magnitude higher than today. In addition, the new trackers need to contribute to the first level trigger in order to maintain a high data-taking efficiency for the interesting processes. Novel detector technologies have to be developed to meet these very challenging goals. 

The German groups active in the upgrades of the ATLAS and CMS tracking systems have formed a collaborative \textquotedblleft Project on Enabling Technologies for Silicon Microstrip Tracking Detectors at the HL-LHC\textquotedblright (PETTL), which was supported by the Helmholtz Alliance \textquotedblleft Physics at the Terascale\textquotedblright during the years 2013 and 2014. The aim of the project was to share experience and to work together on key areas of mutual interest during the R\&D phase of these upgrades. Five work packages have been selected:
\begin{enumerate}
\item[WP1:]{Exchange of Experience}
\item[WP2:]{Radiation Hardness of Silicon Sensors}
\item[WP3:]{Low Mass System Design}
\item[WP4:]{Automated Precision Assembly Procedures}
\item[WP5:]{Irradiations}
\end{enumerate}
Three workshops have been organized in the course of this project: 
\begin{itemize}
\item 28.\,2. - 1.\,3.\,2013 at Mainz University; 
\item 6.\,3. - 7.\,3.\,2014 at Göttingen University; 
\item 1.\,12. - 3.\,12.\,2014 at DESY Hamburg.        
\end{itemize}
The slides of the talks can be found at:\\ 
\href{http://www.terascale.de/research\_topics/rt3\_detector\_technologies/pettl\_project/}{http://www.terascale.de/research\_topics/rt3\_detector\_technologies/pettl\_project/}

This report summarizes the obtained results.

\section{Exchange of Experience}
The current silicon strip detector systems of the two general-purpose LHC experiments ATLAS and CMS differ in many design aspects. The aim of this work package was to exchange the experience that was collected during the different stages of design, construction and running of the detectors in order to identify weak and strong points, for the benefit of the ATLAS and CMS tracker upgrades for HL-LHC. 

\subsection{Lessons learned from design, construction and operation of current LHC tracking detectors\label{section:LessonsLearned}}
This section describes the key findings from the initial half-day session on WP1: {\it{Exchange of Experience}}. This session was held on 28$^{th}$ of February 2013 in the framework of the 6$^{th}$ HGF detector workshop in Mainz, with participation of a large number of international experts from both experiments. The topics covered in the presentations included a detailed and comparative discussion of the module designs realized in the present detectors, integration reports from barrel and end-cap, and overviews of the running experience from the silicon strip systems in both experiments. Below is a brief summary of main recommendations from the speakers, summarized by topic.

\subsubsection{General design issues}
\begin{itemize}
\item Keep it simple.
\item If at all possible, go with \textquotedblleft commercial off-the-shelf solutions\textquotedblright.
\item Aim for a coherent system development, rather than development driven by components. 
%Tim Jones: {\it{Choose Integration by Design rather than Integration by Assembly}}.
\item The foundations of any good design are well-defined sets of requirements and specifications, which are consistent throughout your project. These are also essential to develop meaningful Quality Assurance (QA). 
\item Do not make your production chains any more complicated than necessary. It is already hard enough to deal with the logistics necessary for production in an extended collaboration with different skill sets and equipment. 
\item Reduce the number of part types to a minimum. It is really easy to just add another hybrid, module or cable type to make something else easier -- you will pay for this.
\item Most mechanical problems could have been found earlier. It is essential to stress all components to and beyond their limits prior to the start of mass production.
\item It is necessary to increase the amount of parts available for prototyping, and to handle them in the same way as during production. Put parts into the final environment as soon as possible.
\item \textquotedblleft Babying\textquotedblright rare prototypes puts off finding serious issues. 
\item Production times always get squeezed. Design for peak rates that are 2-3 times higher than original plans. 
\item Design production to allow for scalability. Some CMS facilities had to increase their production rates by a factor of 3.
\item Proper failure management is important. It requires a plan to deal with any potential (critical) fault in the system. Plan the services and how they tie into the system.
\item Use realistic prototyping (adequate scale, using realistic conditions and equipment).
\item If your readout system has a bottleneck in it, it is much better if it is off-detector rather than on-detector.
\item The concept of petals, as compared to mounting modules directly on disks, increases the complexity of the routing of services, increases the number of required module flavours, and adds support material. These drawbacks need to be balanced against potential benefits, such as the possibility to exchange faulty petals and the enabling of a more distributed production. 
\end{itemize}

\subsubsection{Front-end hybrid design}
\begin{itemize}
\item Keep the front-end (FE) hybrids as simple as possible. Use novel materials only when absolutely required. Do not use custom technology from one vendor. Otherwise there is a major risk of getting stuck with single vendors which makes logistics harder than necessary. 
%Tony Affolder: {\it{If you think custom technology is necessary, think again.}} 
\item Plan for volume manufacturability from the start. Do not push the technology to the limit unless really necessary. It makes finding alternate vendors very difficult. 
\item Design FE hybrids with industry from the beginning. Use standard design rules, i.e. ask industry what would be comfortable for production of several ten thousands of parts over 2-4 years. 
\item FE hybrids must be part of the system level design from the start. It is too late to design them once ASICs and/or mechanics are frozen. 
\item Include material where needed, especially ground and power. Over-constraining radiation length makes hybrid production very difficult, or even impossible. Include panel level test structures to reduce QA turn-around times.
\item Be aware of reduced lifetime of smaller feature size technologies. Each new generation of ASICs is wearing out faster than the previous one.
\end{itemize}

\subsubsection{Testing and Quality Assurance}
\begin{itemize}
\item Testing and operation should work under standard conditions.  If non-understood conditions are needed for good operations take this as a strong sign that something is not understood, or not under control. 
\item A system test of relevant size with all the final hardware should be done as early as possible.
\item QA automation is mandatory when large through-puts are required. A database back-end is recommended.
\item Re-check quality at each step (even after a shipment) and track the history of parts.
\item Consider visual inspections, destructive tests on samples and check the quality of signals.
\item Pay attention also to trivial, cheap and low-tech elements.
\end{itemize}

\subsubsection{Module design}
\begin{itemize}
\item Keep the modules as simple as possible. As for hybrids, only use novel materials when necessary. The single vendor risk should be avoided.
\item Tight tolerances need to be justified strongly. Do not pick tolerances just because they are achievable somewhere. They may be difficult to maintain or to propagate to different sites. In the end, they will be relaxed if not achievable and well motivated.
\item When there are many module types, make sure to test carefully all flavours. The simplest case may hide problematic issues or traps.  
\item Avoid many module flavours. Consider that each flavour requires its own gluing jig, bonding jig, transport plate etc. Consider also that in a distributed production scheme all components have to be available at the right place at the right time, which gets much more difficult when there are many different components.
\end{itemize}

\subsubsection{Logistics}
\begin{itemize}
\item Shipments should be avoided (where possible). However, due to the sheer number of modules and the diverse locations of participating institutes it seems inevitable to ship assembled parts. The alternative would be to have the whole production based at CERN, with all manpower and infrastructure located there. This seems unfeasible. 
\item Transport is an important part of integration, requiring thorough preparation. For large units, transporting is a high-risk enterprise. The risks might be so complex that the effort to understand and control them might be unmanageable. 
\item Involve professional shipping companies. They usually know their part of the task well, for them it is just another job.
\item A distributed production implies lengthy knowledge transfer, long learning curves, well-organized logistics and risky, expensive and time-consuming shippings. 
\end{itemize}

\subsubsection{Mounting, services and integration}
\begin{itemize}
\item Keep it simple. Solutions which might look practical on the bench can become very complex, when scaled to the real thing. Think through ahead what you are going to do with the services throughout the assembly. Do optimize your service break points. Prototyping all components needs to be done in a realistic environment (including handling).
\item Plan your services carefully, including planning for their installation. Avoid complex activities with limited access. 
\item Full-scale service mock-ups are essential. Go through the effort to make them real.
\item Avoid safety factor creep. Safety factors should be defined as part of the specifications and then not be added arbitrarily.
\item Keep module mounting (design, tooling, etc.) simple. Manual work by trained operators is often quite adequate. Assembly of modular components which are well tested at previous levels makes the assembly simpler. In this way one can focus on the job itself. 
\item Climbing the production curve requires focused activity and the wider project gets forgotten. People turn into module builders and stave, petal or disk assemblers. The risk is leaving a hole in the "brain-power" devoted to final assembly and sub-system integration.
\item Try to avoid the usage of thermal grease for the mounting of modules, as its usage makes the removal of modules, if repair shall be required, very difficult.
\item Bundle optical fibers into ribbons as early as possible. 
\end{itemize}

\subsubsection{Activation issues}
\begin{itemize}
\item For HL-LHC the activation will be very significant. Taking longer to install the trackers will be a luxury we will just not have.
\item After just a few years access for simple repairs will not be possible - we need to think very hard ahead of time. 
\item Plan for rapid installation and removal of detectors, including remote handling.
\item It is important to qualify models of activation. A risk analysis should be done: have we thought through the consequences of all failures? One should minimize the risk of parts failing by extensive qualification, reliability assurance and system tests.
\end{itemize}

\subsection{Confronting tracker upgrade plans with lessons learned}
In the second PETTL Workshop, held on 6$^{th}$ of March 2014 in G\"ottingen in the framework of the 7$^{th}$ HGF Detector Workshop, WP1 confronted the lessons learned (Section \ref{section:LessonsLearned}) with reality. Are those lessons being taken into account in the Phase-2 Upgrades of the ATLAS and CMS trackers, respectively?

\subsubsection{The case of ATLAS}
The planned Phase-2 Upgrade of the ATLAS Inner Tracker (ITk) is to a large extent driven by the same community as the people contributing to the lessons discussed in this WP.  Therefore, it should be expected that the ITk strip system will incorporate many of the lessons learned described above. The design of the ITk strip system is rather conservative, featuring a barrel section with four strip barrels and two end-cap sections with nine disks each. In contrast to the existing ATLAS strip system, where the barrels were made up from double-sided modules mounted individually on carbon structures, the new barrels will be populated with integrated objects called staves. A stave consists of a carbon fibre core, containing the cooling services and supplying mechanical rigidity, and single-sided modules on both sides. The modules are comparatively simple objects, made by gluing a hybrid directly onto the sensor. A stave-style concept is also applied in the end-caps, where the staves take the form of petals, with modules made from wedge-shaped sensors. The complicated end-cap geometry means that six different sensor types and hence six module types are required. 

The one key rule (to keep things simple) has generally been applied in the design of the system as well as in the modules. The design follows the recommend \textquotedblleft Integration by Design\textquotedblright approach, and uses commercial off-the-shelf (COTS) components and technology wherever feasible. For some items like sensors and read-out ASICs it is evident that COTS solutions do not exist, but this appears inevitable due to the research nature of the project. 
The large number of module types in the forward region contradicts the rule to keep the number of parts to the minimum. However, based on constraints from tracking software, material budget and the general experience of the involved groups, ATLAS choose the petal design with six module types over a rectangular single-type design as foreseen in CMS. We hope to use the lessons from this WP to avoid any related extra complexity, for example in the production logistics, as this has a clear potential to be a bottleneck, based on the CMS experience from building the present CMS tracker.  

Looking at particular items, silicon strip sensors in large numbers have for more than a decade been close to a traditional single-vendor case. However, present ATLAS ITk sensors are designed such that several other vendors could in principle supply them. The process to qualify other sensor vendors is underway. 
The FE hybrids are four-layer flex PCBs using only standard materials, and with track-and-gap and via dimensions that follow a widely accepted industry standard. Therefore, hybrids could be obtained from a large choice of suppliers. This competition on the supplier side in turn means that ITk hybrids are available at stable and moderate cost. 
The design and routing of services has largely been done as a compromise between ease of installation and space constraints. 
The planning for QA strategies and procedures is not yet developed, but is recognized as a vital part and will be designed  also considering the lessons in this very report.

\subsubsection{The case of CMS}
The CMS Tracker upgrade for the HL-LHC will differ substantially from the present CMS tracker, due to the necessity to contribute tracker information to the Level 1 (L1) trigger. Therefore it is not always possible to build on past experience, and to apply the lessons learned from construction of the present tracker. New techniques have to be learned, and the rule to keep things simple cannot always be applied. In addition, the CMS Tracker upgrade is still in the R\&D stage and many lessons concerning mass production, QA and related areas are not yet relevant. 

The baseline layout of the tracker upgrade is a classical barrel plus end-cap design, which was optimized with a custom software tool, considering resolution, hermeticity, material budget, cost and many other aspects. The outer barrel comprises three cylindrical layers and is constructed from longitudinal structures (referred to as rods), each with 12 silicon modules. In the inner barrel the number of modules on the three cylindrical layers are optimized individually. In the end-caps five disks are forseen per side, each disk equipped with 15 rings of modules. There are 15\,348 modules in total. The overall layout can be considered \textquotedblleft standard\textquotedblright, and the fact that all disks are identical poses a simplification with respect to the present detector, where four types of disks were used.

CMS considers also an alternative layout in which the inner barrel is constructed differently. In this approach the modules are progressively tilted such that the sensor surface is approximately perpendicular to particles from the primary interaction point. Motivations are a reduction of the number of required modules, and consequently of the power, by 30\,\%, a 5\,\% lower cost and a reduction of material in the region around a pseudorapidity\footnote{ Both ATLAS and CMS use a right-handed coordinate system, with the origin at the nominal interaction
point, the $x$-axis pointing to the centre of the LHC ring, the $y$-axis pointing up (perpendicular to the LHC plane), and the $z$-axis along the anticlockwise-beam direction. Cylindrical coordinates ($r$, $\phi$) are used in the transverse plane, $\phi$ being the azimuthal angle
around the $z$-axis. The polar angle $\theta$ is measured from the positive $z$-axis. The pseudorapidity $\eta$ is defined as $-\ln{[\tan{(\theta/2)}]}$.} of $\pm 2$. This spectacular design does not follow the \textquotedblleft keep it simple\textquotedblright rule. The routing of services, mentioned in the lessons as one of the key troublesome areas, will be more complicated than in the classical approach. The routing of services has not yet been studied in detail.

The new CMS tracker concept forsees sending out trigger data at 40\,MHz, plus sending out the full read-out data upon reception of a L1 trigger. The former is possible due to on-detector data reduction. Uninteresting low transverse momentum tracks are rejected by exploiting their bending in the 3.8\,T magnetic field of the CMS solenoid. The silicon modules, referred to as \textquotedblleft $p_T$ modules\textquotedblright, consist of two closely spaced sensor layers. Hits in those layers are associated locally and two-hit tracklets (\textquotedblleft stubs\textquotedblright) are formed. The distance of the two hits, expected to be small for stiff high $p_T$ tracks, is evaluated and only stubs consistent with a $p_T >$\,2\,GeV are sent out. Tracks are formed from the stubs at the back-end. This new concept drives the design of the new modules, front-end ASICs and hybrids. 

Two basic modules types are used: modules with two strip sensors, referred to as 2S $p_T$ modules, and modules with a strip and a macro-pixel sensor, referred to as PS $p_T$ modules. \\
The 2S modules are used for radii above 60\,cm, i.e. in the outer barrel and the outer part of the disks. Two identical 10\,cm $\times$ 10\,cm large sensors with 5\,cm long strips of 90\,$\micro\meter$ pitch are used. The sensors are wire-bonded to the hybrid both from the top and from the bottom. 

The PS modules, aimed to provide also $z$ information for the barrel part of the track trigger, are used in the inner barrel and the inner part of the disks. The sensor area amounts to 5\,cm x 10\,cm. The strip sensor comprises strips with 2.5\,cm length and the macro-pixels in the other sensor are 1.5\,mm long. The pitch is 100\,$\micro\meter$ in both cases. 

As a consequence of the usage of two different module types two read-out ASICs need to be developed: the CMS Binary Chip (CBC) for the read-out of the strip sensors, and the Macro-Pixel ASIC (MPA) for the read-out of the pixel sensor. The latter chip is bump-bonded to the sensor. The unavoidable necessity of two read-out chips poses a considerable increase of required effort with respect to the present read-out electronics, where the APV25 ASIC was used to read out the strip modules in the whole tracker. 

While two basic module types are required, the total number of module flavours has been reduced as compared to the present detector, where nine basic module geometries were used (seven of which were required in the end-caps). In total there were 15 different sensor designs and twelve different hybrid designs, leading to a zoo of 29 different module designs. The CMS experience with this high number of module flavours was rather negative: prototyping was lengthy, the complexity for cooling and mounting was enormous, the number of jigs exploded and the logistics was a nightmare. Based on this lesson there will be only five module variants: two for the 2S modules and three for the PS modules. Variants of the same module differ only in the sensor spacing, and it is believed that the differences are thus relatively local (e.g. the sensors are the same). This reduction of module types is made possible through the decision to use rectangular modules also in the end-caps, while wedge-shaped modules were used in the past. This comes at the cost of a 5-10\,\% increase in material budget, considered acceptable, and has also the consequence that petals cannot be used anymore, since their wedge-shaped area can hardly be covered by rectangular modules. The overall hope is that module prototyping and module production can be much more stream-lined as compared to the past effort. 

Looking now more into the details of the three sub-detectors, the outer barrel design is basically a duplication of the present design utilizing the concepts of wheels and rods. Differences include tighter dimensions and a different location of the cooling pipes (now routed inside the rods as compared to outside), but overall the project looks doable and \textquotedblleft easy\textquotedblright, judging from the success of the present outer barrel.

The design of the inner barrel is more complicated and less advanced as the one of the outer barrel, and differs more from the present design. Previously modules were mounted on longitudinal \textquotedblleft strings\textquotedblright, while now they will be placed directly onto the cylinders. Flaws of the previous system (crowded service arrangement, detaching of cooling contacts) are well known. At this point the design of the inner barrel is not yet at a level that would allow to judge potential improvements. In particular the routing of the services has not yet been worked out, and the decision between a cylindrical and a tilted geometry, which will have considerable consequences for the service routing, has not yet been made.

From all sub-systems, the end-caps differ most from the present design. Based on the lessons learned, but also due to the decision to use rectangular modules, the petal concept was abandoned. Instead modules will be mounted directly onto the disks. This is believed to reduce the amount of material. Mechanically the new approach is not as simple as expected, as clashes between the rectangular modules can occur. The increased simplicity in one area (less module flavours) is paid by additional complications elsewhere. The routing of services is already known to be difficult, as no free radial corridors remain on the disks, and fibers, for example, might have to be routed on top of the modules.

The 2S $p_T$ modules consist of a lower and an upper sensor, glued to AlCF spacers that provide the mechanical distance and act as cooling contacts. Each sensor is wire-bonded on both sides (top and bottom) to FE hybrids that carry eight CBC read-out chips each. A service hybrid carries the DC-DC converters and electronics for electro-optical conversion, and provides the bias contact. The module design is still being iterated. Due to this and due to lack of components, no full-size prototypes - considered very important in the lessons learned - do exist yet, although smaller prototypes have already been developed (e.g. a module with two CBCs). One of the lessons was to use as few exotic materials as possible. This is followed, with the only \textquotedblleft exotic\textquotedblright material being AlCF, used for the spacers. The thermal and mechanical performances are being studied with finite element (FE) simulations and were design drivers from the beginning. The critical thermal performance of several early designs could be shown, and improvements to the design were identified. The verification of those simulations with measurements on real modules or at least thermal mock-ups is still pending, which points again to the importance of early prototyping. 

Another reoccuring theme in the lessons learned was the importance of FE hybrid development and production, based on past negative experiences. CMS has focussed very early on this topic. The FE hybrids are very challenging: they receive signals from the top and the bottom, and thus have to accomodate a large number of channels, which necessitates flip-chip assembly of the ROCs. The trace width and spacing amounts to $25\,\micro\meter$ and laser vias with a diameter below $50\,\micro\meter$ are needed. The number of potential vendors is small, something that was seen critical in the lessons. As requested in the lessons, industry was involved from the very beginning, and in a very systematic way. Both rigid and flex versions were tried. The rigid version was supposed to be bondable from top and bottom, while the thinner flexible version has to be folded around a stiffener, to allow for wire-bonding from both sides. The 6-layer rigid version was very thin ($265\,\micro\meter$) and has been abandoned due to its flexibility, leading to wire-bonding problems. The 4-layer flex version seems to be a viable solution. Nevertheless, production issues were encountered also there. This project is being followed closely, to limit the risk for future production and yield issues. 

The PS $p_T$ modules received attention later, as they were expected to profit from the work on the 2S modules. However, PS modules are more complicated, and pose their own challenges. In particular the heat load is larger, and the macro-pixel sensor needs to be cooled through the ASIC layer. Again, FE simulations helped to identify issues and to improve the thermal management. Also here one of the problems is the lack of components, required for prototyping. This is now being adressed: since the full MPA will not exist for a while, a small version (MPA-light) with fewer pixels is being developed. The assembly will be tested in the Macro-Pixel SubAssembly Project (MaPSA-light), where a MPA-light will be bump-bonded to a small sensor. A full-scale MaPSA is presently not expected before 2016. The clear development plan allows to move forward, although there is a risk to push crucial but complex aspects (stub finding logic, through-silicon vias) to the future. 

In summary, in areas where CMS suffered most in the past, namely the number of module flavours and the issues encountered during FE hybrid development, lessons have clearly been learned. Module flavours have been reduced through conceptional changes, and FE hybrid development has been persued very systematically from the start. The decision to move from the petal concept to a disk concept was partly following from the decision to use rectangular modules also in the end-caps, partly it was motivated by past experience. The \textquotedblleft keep it simple\textquotedblright lesson is not always followed, sometimes because the new track trigger concept does not allow for it, sometimes because simplicity is balanced against other aspects such as cost or material budget. The usefulness of the early construction of prototypes or at least mock-ups is acknowledged, but due to the lack of components simulations often have to be used to identify problems and to guide the design. Nevertheless a clear effort is made in order to improve this situation.

\section{Radiation Hardness of Silicon Sensors}
\begin{figure}%
		\centering%
		\includegraphics[width=0.59\linewidth]{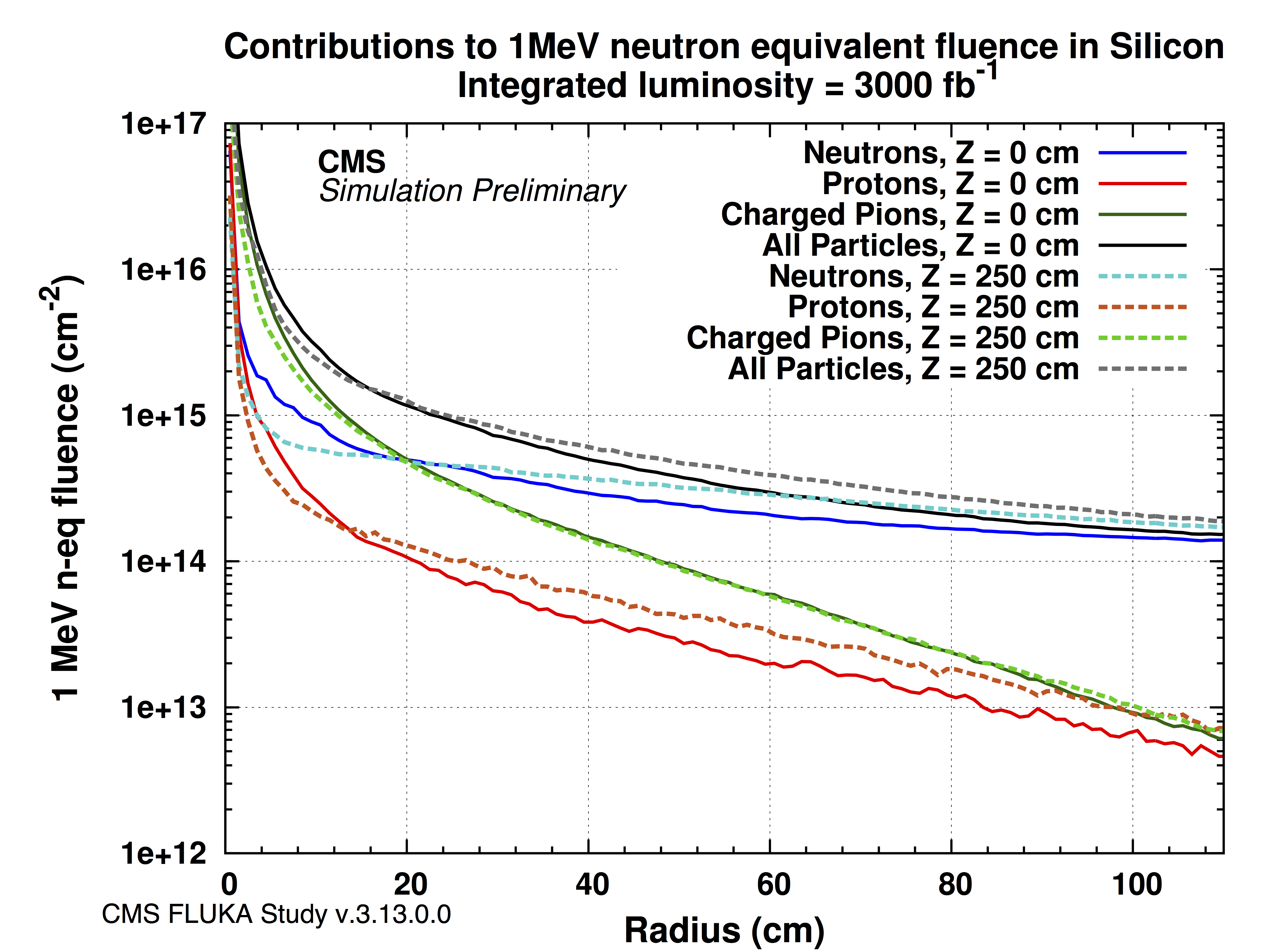}%
		\caption{The expected 1\,MeV neutron equivalent fluence in silicon in the CMS Tracker, simulated with FLUKA and plotted as function of the radial distance from the (nominal) beam line for two values of $z$, namely $z = 0$ (solid lines) and $z=250$\,cm (dashed lines)\,\cite{Fluka}. The total fluence, as well as the individual contributions from neutrons, protons and charged pions, are shown for an integrated luminosity of $3000$\,fb$^{-1}$.}%
		\label{fig:slhcfluence}%
\end{figure}

The detector performance of the HL-LHC experiments will degrade during their runtime as a result of radiation damage.
Figure \ref{fig:slhcfluence} demonstrates the expected fluence after 10 years of operation in the CMS experiment. A maximum 1\,MeV neutron equivalent fluence, $\Phi_{eq}$, of approximately $2\times 10^{15}$\,n$_{eq}$cm$^{-2}$ is expected for the strip region of the detector. Comparable radiation damage is expected for the ATLAS experiment, thus the requirements for the sensors in both experiments are fairly similar.

The origin of the most relevant effects are electrically active defects in the silicon bulk, created by non-ionising energy loss. The main consequences are:%
\begin{itemize}%
	\item Change of the electric field in the sensor. This can partially be overcome by higher bias voltages.%
	\item Increase of the leakage current that leads to an increase of the heat-generation and power consumption, and to an increase of the noise. This can partially be overcome by cooling.%
	\item Charge carriers are trapped by defects, leading to lower signals and thus to lower efficiency and a degraded resolution. This can partially be overcome by higher bias voltages.%
\end{itemize}%
In addition to bulk damage, ionising energy loss leads to surface damage. The main consequences are high fields close to the surface, a degradation of the signal near to the surface, an increase of the surface current and a reduction of the breakdown voltage in p-in-n devices\,~\cite{sch14a}. 

In order to choose the optimal sensor parameters (such as material, polarity, bulk thickness and surface design) and to improve the radiation hardness of silicon sensors, both experiments have initiated extensive sensor and irradiation campaigns. Within these campaigns several approaches have been followed. 
Electrical parameters and signal generation in silicon test sensors have been measured extensively before and after irradiation. This resulted in a comprehensive understanding of radiation damage effects in silicon and established a baseline for the sensor production for both experiments (the strip sensor region in the ATLAS experiment and the outer tracker for CMS). In the following, an investigation of the inter-strip isolation is presented in Sect. \ref{sec:interstrip}, followed by results on surface damage in Sect. \ref{sec:surf}. A summary on trapping parameters is given in Sect. \ref{sec:trap}. Results from ATLAS and CMS on the leakage current increase are compared in Sect. \ref{sec:leakage}, while a comparison of the Charge Collection Efficiency (CCE) in strip sensors is presented in Sect. \ref{sec:cce}.

The impact of the irradiation can be modeled with TCAD finite element programs. Simulations have been established as a major tool to qualitatively understand radiation damage effects in silicon. Moreover, by developing defect models for these simulations it is possible to predict electrical parameters and the charge collection of sensors as a function of the fluence. The accuracy of these predictions is strongly related to the accuracy of the underlying defect model and the understanding of radiation damage in silicon. Section \ref{sec:models} summarizes the main outcome of the simulation effort. A better understanding of radiation damage and an improvement of radiation models is envisaged by investigations of defects in the silicon bulk. An overview of the recent results is presented in Sect. \ref{sec:defects}.

\subsection{Inter-strip isolation}
\label{sec:interstrip}
Segmented n-in-p type sensors need additional process steps in order to establish isolation between the n-doped structures. This is necessary due to the formation of an accumulation layer of electrons which are attracted towards positive charge in the coupling oxide. An isolation layer in form of p$^{+}$ dopands between the collecting n$^{+}$ implants prevents an undesired charge sharing over several strips if the p$^{+}$ doping concentration is high enough. On the other hand, the electrical characteristics like leakage current and breakdown voltage strongly depend on the concentration as well. Hence the calculation of the p$^{+}$ concentration is a critical point in order to develop radiation hard n-in-p type silicon sensors. 

\begin{figure}
		\centering
		\includegraphics[width=0.9\linewidth]{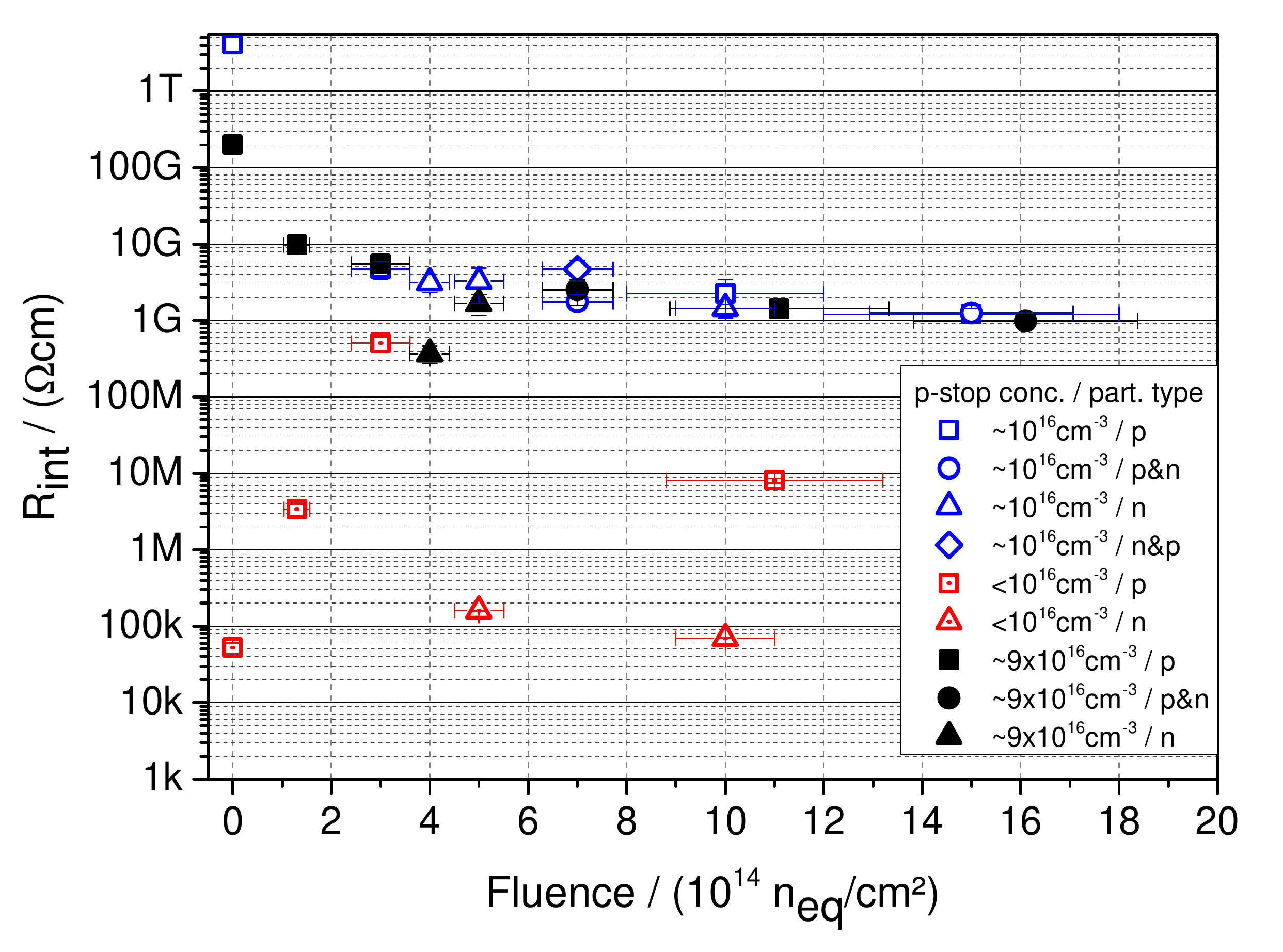}
		\caption{Inter-strip resistance for three different p$^{+}$ doping concentrations as a function of the fluence. Measurements have been performed at a bias voltage of $V_{bias}=-600$\,V and at a temperature of $T=253$\,K. A voltage was applied to the neighbouring DC pads and ramped up to 1\,V, while the potential of the measured strip was kept at the potential of the bias ring. The notation \textquotedblleft n\textquotedblright, \textquotedblleft p\textquotedblright, \textquotedblleft n\&p\textquotedblright and \textquotedblleft p\&n\textquotedblright in the legend refers to the irradiation: \textquotedblleft n\textquotedblright and \textquotedblleft p\textquotedblright stand for irradiation with neutrons and protons, respectively, while \textquotedblleft n\&p\textquotedblright and \textquotedblleft p\&n\textquotedblright mean mixed irradiation, with first neutrons and later protons, or vice versa, respectively. The fractions of neutron and proton fluences in mixed irradiations were chosen according to the expectation (Fig.~\,\ref{fig:slhcfluence}).}
		\label{rint}
\end{figure}

In this study sensors from three different vendors with three different initial doping concentrations of an atoll-shaped p-stop isolation layer have been 
studied. The inter-strip resistance is an indicator for the quality of the strip isolation and has been measured before and after irradiation with protons and neutrons. The aim of the study is to determine a suitable doping concentration which ensures a sufficient sensor performance for the expected lifetime of about 3000~\SIunits{\femto \reciprocal \barn}.
Figure \ref{rint} shows the evolution of the inter-strip resistance with the expected fluence for the HL-LHC. Samples from Hamamatsu Photonics K. K. (HPK)\footnote{http://www.hamamatsu.com} and CNM\footnote{http://www.imb-cnm.csic.es} with p$^{+}$ doping concentrations of about \SIunits{$1\times$\power{10}{16}}{\,\cm \rpcubed} and  \SIunits{$9\times$\power{10}{16}}{\,\cm\rpcubed} show after irradiation an inter-strip resistance higher than \SIunits{200\,}{\mega\ohm} for fluences up to \SIunits{$1.6\times$\power{10}{15}}{\,\cm \rpcubed}, which is high enough to ensure sufficient strip isolation. No difference in the reduction of the inter-strip isolation after irradiations with protons with respect to irradiations with neutrons is observed. Although the samples with a peak doping concentration of \SIunits{$9\times$\power{10}{16}}{\,\cm \rpcubed} show a sufficient inter-strip resistance, they also exhibit high strip noise after irradiation. This noise can be related to the high p$^{+}$ doping concentration as very high electric fields arise at the p$^{+}$ implant edges. Samples produced by ITE\footnote{http://www.ite.waw.pl} with a doping concentration of significantly less than \SIunits{$1\times$\power{10}{16}}{\,\cm \rpcubed} cannot ensure sufficient strip isolation, with inter-strip resistances in the range of few \SIunits{\mega \ohm} and less.

These measurements indicate that the peak doping concentration for the atoll-shaped p-stop isolation has to be tuned carefully and an optimal value might be around \SIunits{$1\times$\power{10}{16}}{\,\cm \rpcubed} and below \SIunits{$9\times$\power{10}{16}}{\,\cm \rpcubed}. 

\subsection[Surface damage]{Impact of surface damage on the sensor performance}
\label{sec:surf}
Surface damage has not been considered as an important source of the sensor degradation for a long time. However, recent studies for p-bulk sensors \cite{erf14} demonstrated that even after low doses of electron irradiation a charge build-up at the Si-SiO$_{2}$ interface can lead to an increase of charge sharing and a decrease of the charge collected in the strip with the highest signal in a signal cluster (seed strip). Depending on the strip and sensor layout and the strip isolation technique, the effect on the seed signal can reach 20\,\%. For hadron-irradiated sensors the degradation is less severe; for a fluence of $2.1\times 10^{15}$\,n$_{eq}$cm$^{-2}$ an effect of 5\,\% was observed.

\begin{figure}
		\centering
		\includegraphics[width=0.98\linewidth]{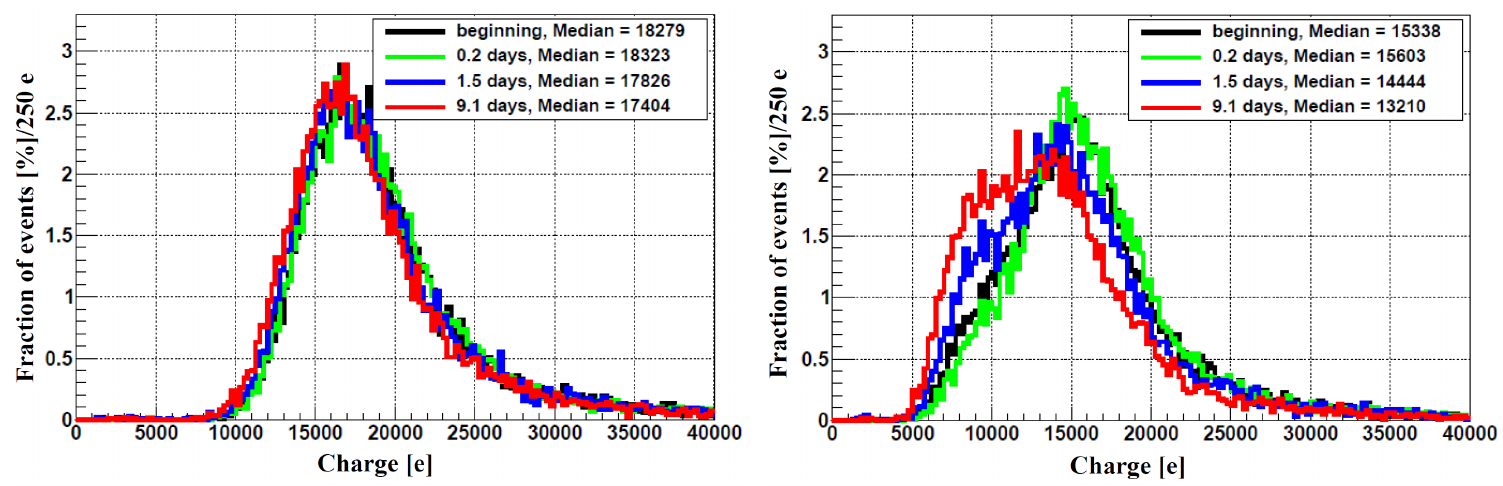}
		\caption{Left: pulse-height distributions for clusters of four strips for a non-hadron-irradiated p-bulk MCz sensor with \SIunits{80}\,{\micro \meter} pitch and p-stop strip isolation, biased at 600\,V. The sensor was exposed to $\beta$ particle doses of 0\,Gy (beginning), 10\,Gy (0.2 days), 75\,Gy (1.5 days) and 450\,Gy (9.1 days). Right: the corresponding pulse-height distributions for the seed strip \cite{erf14, kla14}.}
		\label{fig:surf01}
\end{figure}

Figure\,\ref{fig:surf01} presents the cluster pulse-height distributions for a Magnetic Czochralski (MCz) sensor for different exposure times to a 100 MBq $^{90}$Sr $\beta$ source (left) and corresponding pulse-height distributions of the seed strip (right). The distribution of the full cluster charge demonstrates no obvious distortion resulting from irradiation. In comparison, the distribution of the seed strip is strongly distorted due to charge sharing. Data acquisition was performed with an ALiBaVa multi-channel read-out system at $-20\SIunits{\degree}$C. More information about the setup, data acquisition and analysis can be found in \cite{erf14}. 

\begin{figure}%
		\centering%
		\includegraphics[width=0.43\linewidth]{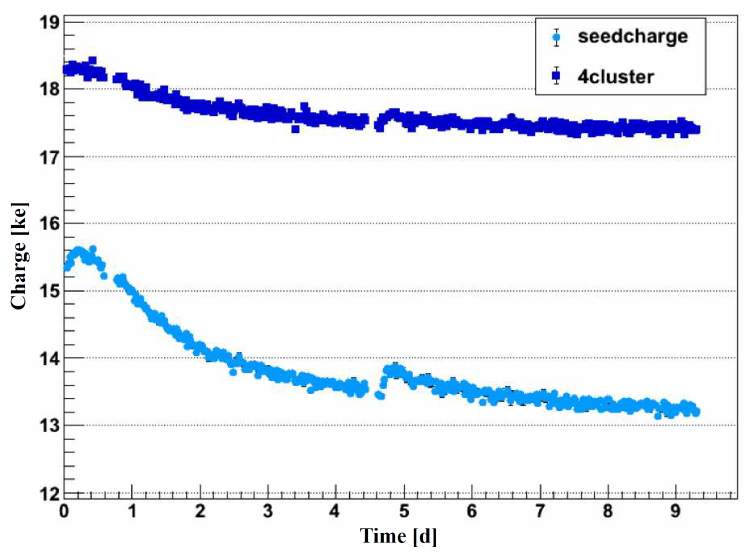}%
		\hfill
		\includegraphics[width=0.47\linewidth]{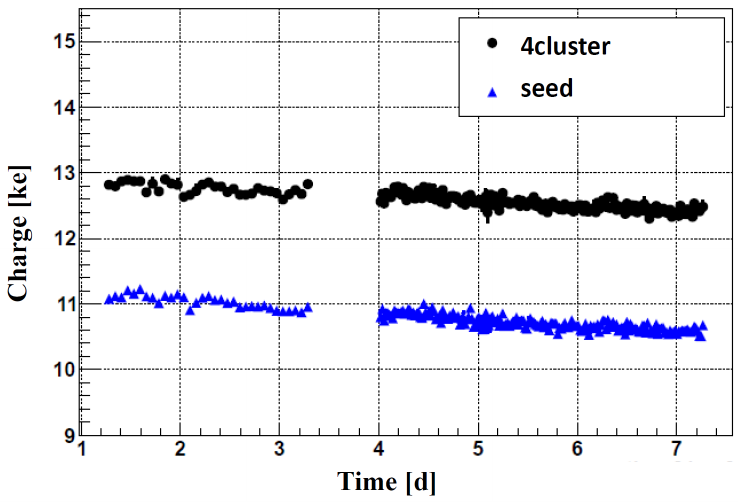}%
		\caption{Left: median of the pulse-height distributions for a cluster of four strips (\textquotedblleft 4cluster\textquotedblright) and the seed strip (\textquotedblleft seedcharge\textquotedblright) as a function of the exposure time to a $\beta$ source for a non-hadron-irradiated p-bulk MCz sensor, biased at 600\,V. Right: corresponding distribution for a hadron-irradiated p-bulk FZ sensor, biased at 1000\,V. Both sensors have a pitch of \SIunits{80}\,{\micro \meter} and p-stop strip isolation \cite{erf14, kla14}.}
		\label{fig:surf02}%
\end{figure}

Figure \ref{fig:surf02}\,(left) demonstrates the reduction of the median of the pulse-height distributions for a cluster of four strips and the seed strip as a function of the exposure time to the $^{90}$Sr $\beta$ source. The time corresponds to the doses described in Figure \ref{fig:surf01}. A degradation of the seed charge by 20\,\% is observed. 
 The p-bulk float-zone (FZ) sensor presented on the right-hand side of Fig.\,\ref{fig:surf02} had been irradiated with 23\,GeV protons to a fluence of $\Phi=$\,\SIunits{$1.5\times$\power{10}{15}}{\,\cm \rpsquared} and with reactor neutrons to $\Phi=$\,\SIunits{$6\times$\power{10}{14}}{\,\cm \rpsquared}. The reduction of the median of the pulse-height distributions for a cluster of four strips and the seed strip is presented as a function of the same exposure time to the $^{90}$Sr $\beta$ source. The reduction due to additional surface damage is about 5\,\% for both, the cluster charge over four strips and the seed charge.
 
Although the seed-charge reduction is lower for hadron-irradiated sensors compared to non-irradiated sensors, the impact of surface damage on thin sensors with a binary read-out is not negligible. Recent simulations with TCAD include oxide charges \cite{sch14a, cms14}, and moreover, there is an ongoing effort to implement interface states to the defect models. 

\subsection{Simulation of trapping rates}
\label{sec:trap}
One of the major issues for detectors at the HL-LHC is the degradation of the resolution and efficiency due to radiation-induced trapping. Recent studies have indicated that the effective trapping rates for holes and electrons are about 50\,\% lower than what was expected from extrapolations of earlier studies performed at low fluences.

The effects of bulk damage from high-energy proton irradiations in silicon is in particular puzzling, because of the mixture between point defects and cluster defects. However, utilizing effective parameters (such as trapping rates and effective trap models) a description of irradiated sensors is possible in the fluence range expected for the strip region of the tracker detectors. Without claiming full understanding of the \textit{real} trapping mechanism that might depend on several effects like local variations of the electric field, defect types and occupation, and others, it is possible to generate an effective model for trapping rates which can be used for detector simulations. 

The PixelAV \cite{pixAV} program, combined with electric field simulations from TCAD (based on defect models, Sect. \ref{sec:models}), can be used to simulate transient current technique (TCT) pulses for electrons and holes. The methodology of the simulation and details about the TCT setup and the measurement procedure used for the following studies are described in \cite{poe14}. The effective trapping rate can be extracted from the simulation by tuning the trapping rate to make the simulated TCT curve fit the corresponding measured TCT pulse. 

\begin{figure}%
		\centering
		\includegraphics[width=1.0\linewidth]{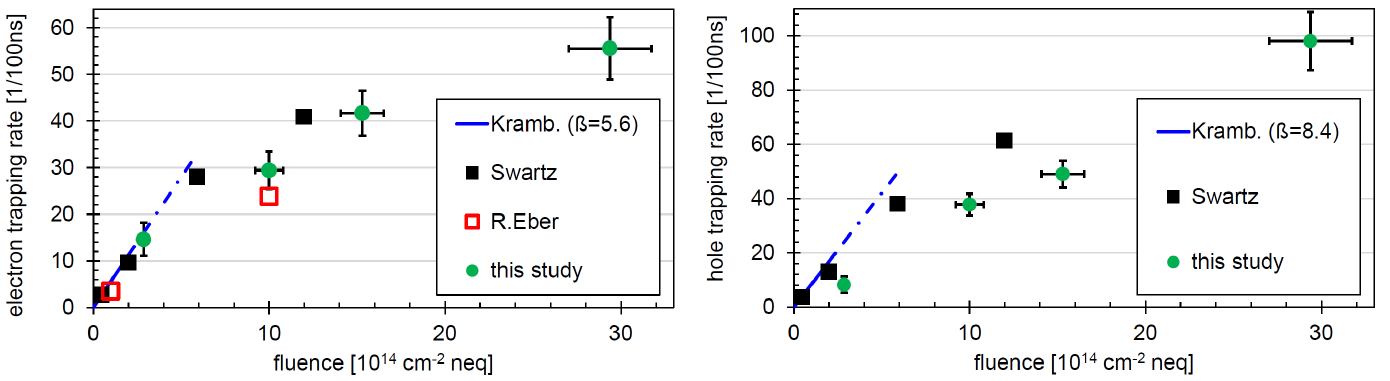}%
		\caption{Effective trapping rates for electrons (left) and holes (right) extracted from 23\,GeV proton irradiated pad-diodes~\cite{poe14}. Literature values are given for reference.}%
		\label{fig:trap3}%
\end{figure}

The extracted trapping rates for high energy (23\,GeV) proton-irradiated sensors are presented in Figure \ref{fig:trap3} for electrons (left) and holes (right). For comparison the earlier measurements at lower fluences (Kramberger, Swartz) and the extracted trapping rates for 23\,MeV proton irradiation (R. Eber) are also shown. The effect of radiation induced trapping is 50\,\% smaller than expected from earlier extrapolations from irradiations with lower fluences. However, trapping still remains the major effect for sensor degradation at the HL-LHC.

\subsection{Comparison between ATLAS and CMS results}
\label{seccomp}
ATLAS and CMS have similar specifications for their silicon strip sensors. This resulted in a common market survey, which is the first step of the CERN procurement process towards a contract with a supplier. The ATLAS and CMS sensor working groups have conducted several projects to find the most suitable silicon material and design for their upgraded trackers. In this section some of these complementing results are discussed.\\%
\subsubsection{Leakage current}
\label{sec:leakage}
The power consumption and heat production of the silicon sensors is highly relevant for the operation of the tracker in both experiments. Due to irradiation the leakage current increases dramatically; in turn this requires efficient cooling of the sensors. However, there are limitations due to the design of the detector modules and the size of the sensors. Therefore it is important to have an expectation of the current at the operation temperature at the highest expected fluence.

\begin{figure}%
		\centering
		\includegraphics[width=0.45\linewidth]{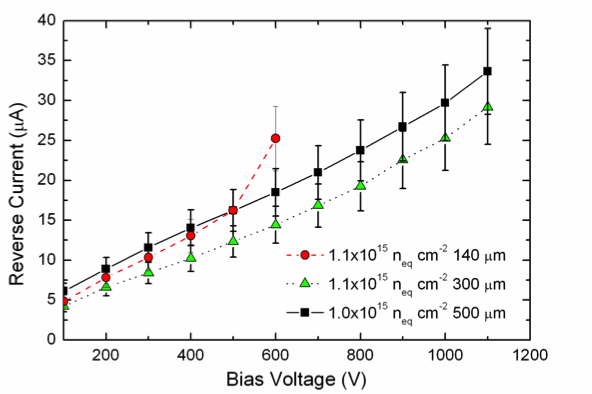}%
		\includegraphics[width=0.48\linewidth]{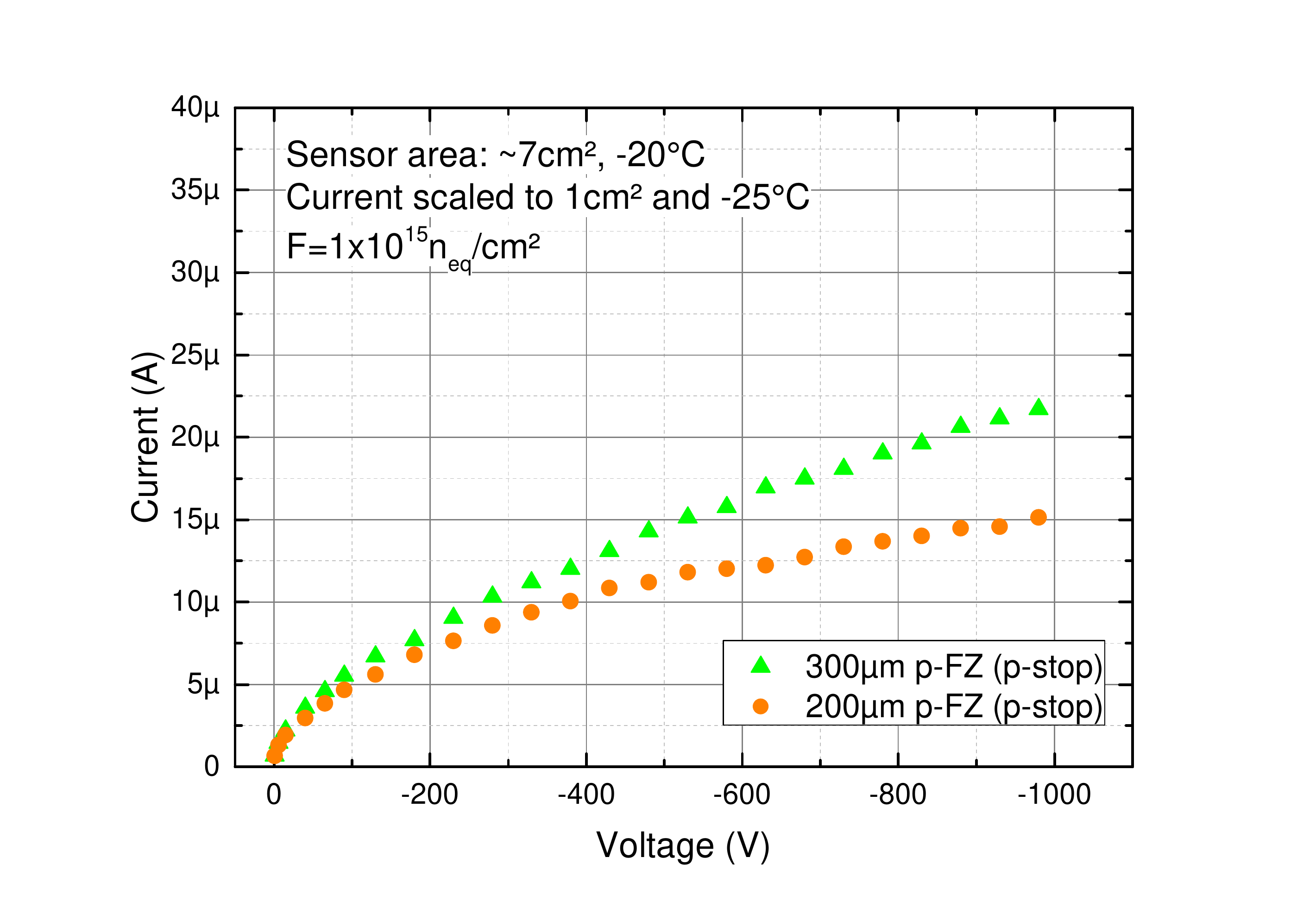}%
		\caption{Current generated in strip sensors after irradiation with $\Phi_{eq}\approx 1\times 10^{15}$\,n$_{eq}$\,cm$^{-2}$ for HPK sensors produced for ATLAS \cite{Affolder} (left) and CMS (right). The abbreviation FZ refers to float-zone sensors.}%
		\label{fig:lc1}%
\end{figure}%

A parameterization of the current as a function of the fluence and annealing time, which is generally used for calculations, can be found in \cite{mol99}. An important observation in pad-diodes is the linear dependence of the current on the sensor volume up to a fluence of $\Phi_{eq}\approx 1\times 10^{15}$\,n$_{eq}$\,cm$^{-2}$. Thinner sensors should therefore demonstrate lower leakage currents, since the volume in which the current is generated is smaller.

Both ATLAS and CMS have further investigated the generation of leakage current in strip sensors. Examples of strip sensor measurements for ATLAS and CMS are given in Figure \ref{fig:lc1}. The sensors produced for the ATLAS experiment (left) have a sensor area of $\sim$~1\,cm$^{2}$. 
There is no significant dependence of the leakage current on the sensor thickness for the thinner devices and also the current for the  \SIunits{500 }{\micro \meter} thick sample does not scale with the thickness as expected. This contradicts the previous observation on pad-diodes. The sensors produced for CMS (right) have an area of $\sim$~7\,cm$^{2}$. For those sensors the currents for \SIunits{320\,}{\micro \meter} and \SIunits{200\,}{\micro \meter} thickness demonstrate the expected dependence on the sensor volume, in contrast to the results presented by ATLAS.

\begin{figure}%
		\centering
		\includegraphics[width=0.6\linewidth]{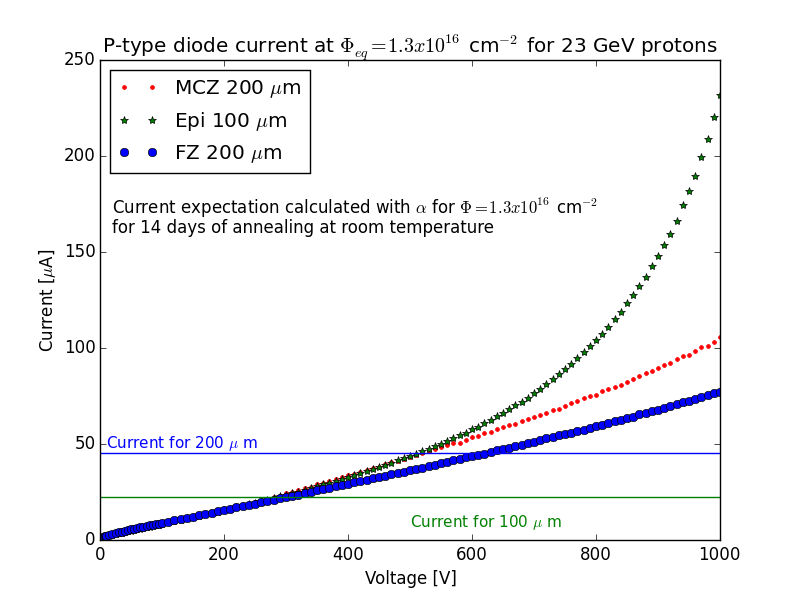}%
		\caption{Current generated in CMS HPK pad-diodes. Float-zone (FZ), Magnetic Czochralski (MCZ) and epitaxially grown (Epi) sensors are compared. The sensors were irradiated with a fluence of $\Phi_{eq}\approx 1.3\times 10^{16}$\,n$_{eq}$\,cm$^{-2}$.}%
		\label{fig:lc2}%
\end{figure}

This behaviour could be explained by the charge multiplication effect. For sensors with very high electrical fields below the readout electrode, impact ionisation can multiply the signal and the leakage current. This effect has been studied by the CERN RD50 Collaboration~\,\cite{rd50} for highly irradiated sensors. The multiplication region is a small region below the electrode, depending on the electric field strength and shape. Therefore the amount of multiplication does not depend too much on the actual volume of the sensor, but on the higher fields in sensors. CMS has seen the effect of charge multiplication in highly irradiated thin pad-diodes (Figure~\,\ref{fig:lc2}). The current generated in those sensors demonstrates no dependence on the volume of the devices and exceeds the expected value. 

\subsubsection{Charge collection studies}
\label{sec:cce}
Considering the expected harsh radiation environment for the silicon sensors, one of the key parameters is the signal size or charge collection, which decreases with fluence. In this section, the cluster charge is used to characterize the charge collection efficiency, where a cluster is formed from a seed strip plus its neighbours. In the case of CMS, for example, the seed strip has to have a signal of at least five times its noise, while the charge of a neighbouring strip must amount to at least twice its noise.

\begin{figure}%
\centering
\includegraphics[width=0.5\columnwidth]{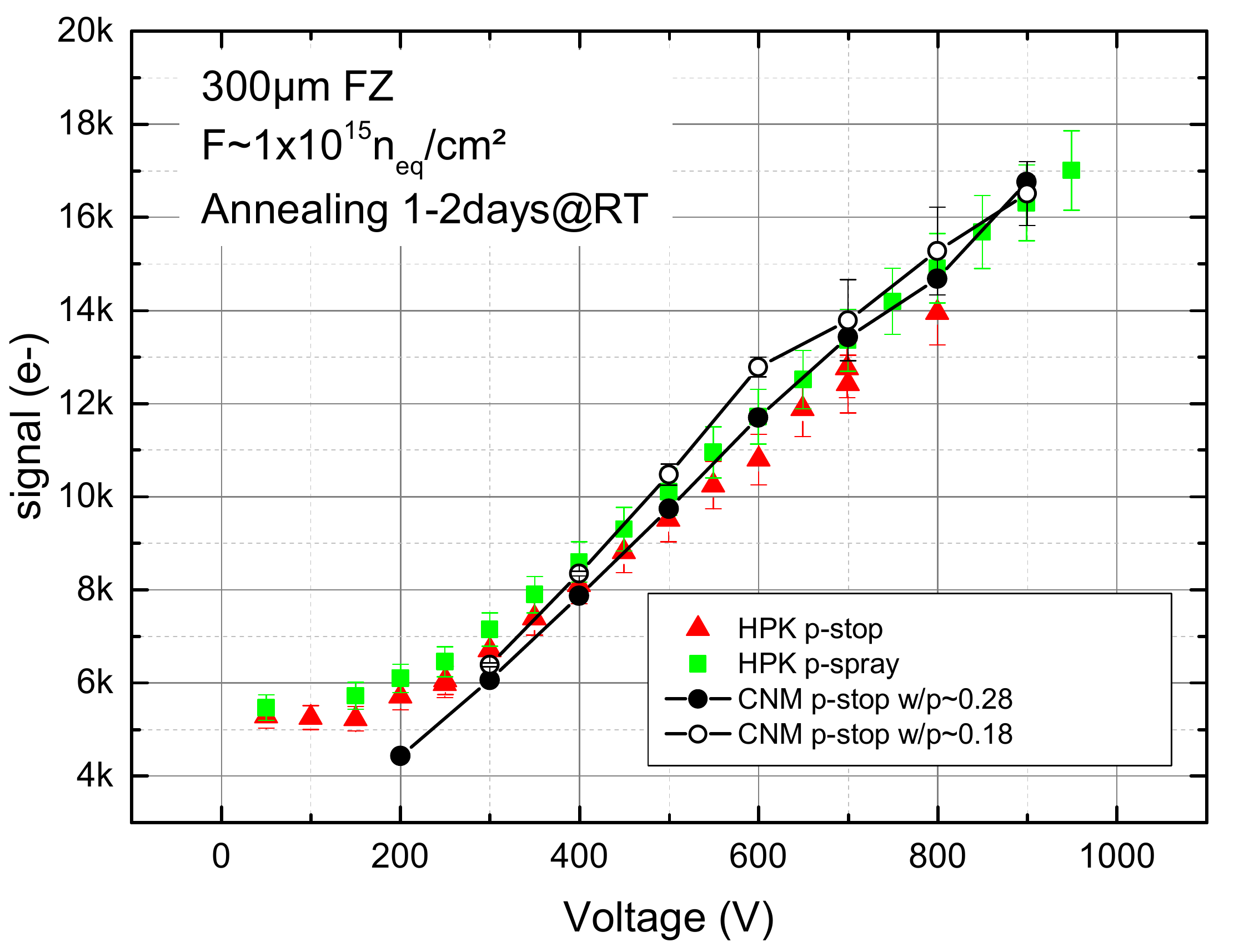}%
\caption{Charge collection after proton irradiation to $1\times 10^{15}$~n$_{\mathrm{eq}}$~cm$^{-2}$, after 1-2 days of annealing at room temperature (RT). The strip sensors have been processed by two different vendors and the strip isolation as well as the strip layout was different (w/p stands for width over pitch). Still, the charge collection does not differ significantly.}%
\label{fig:CC_process_Dierlamm}%
\end{figure}

A comparison was made of p-in-n type FZ sensors from several manufactures (HPK, Micron, CNM) with p-stop or p-spray strip isolation and several strip pitches (75\,$\micro$m - 100\,$\micro$m). The charge collection measurements did not show significant differences between these materials, which demonstrates that the process details do not significantly influence the charge collection in the relevant fluence ($\Phi_{eq} \leq 2\times 10^{15}\,$n$_{eq}$\,cm$^{-2}$) and bias voltage ($<$\,1000\,V) ranges (Fig.~\ref{fig:CC_process_Dierlamm}). 

\begin{figure}%
\centering
\includegraphics[width=0.45\columnwidth]{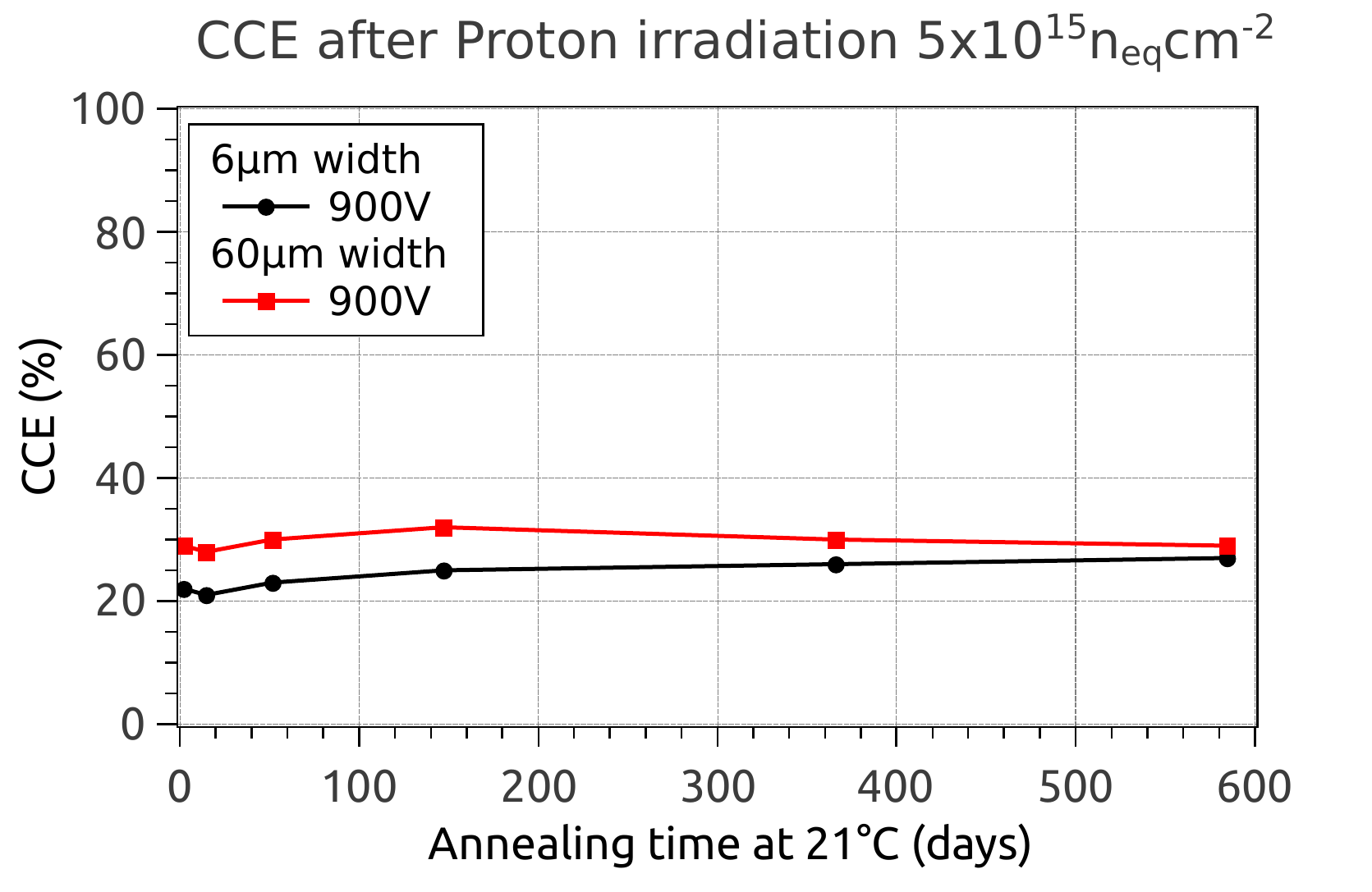}%
\hfill
\includegraphics[width=0.45\columnwidth]{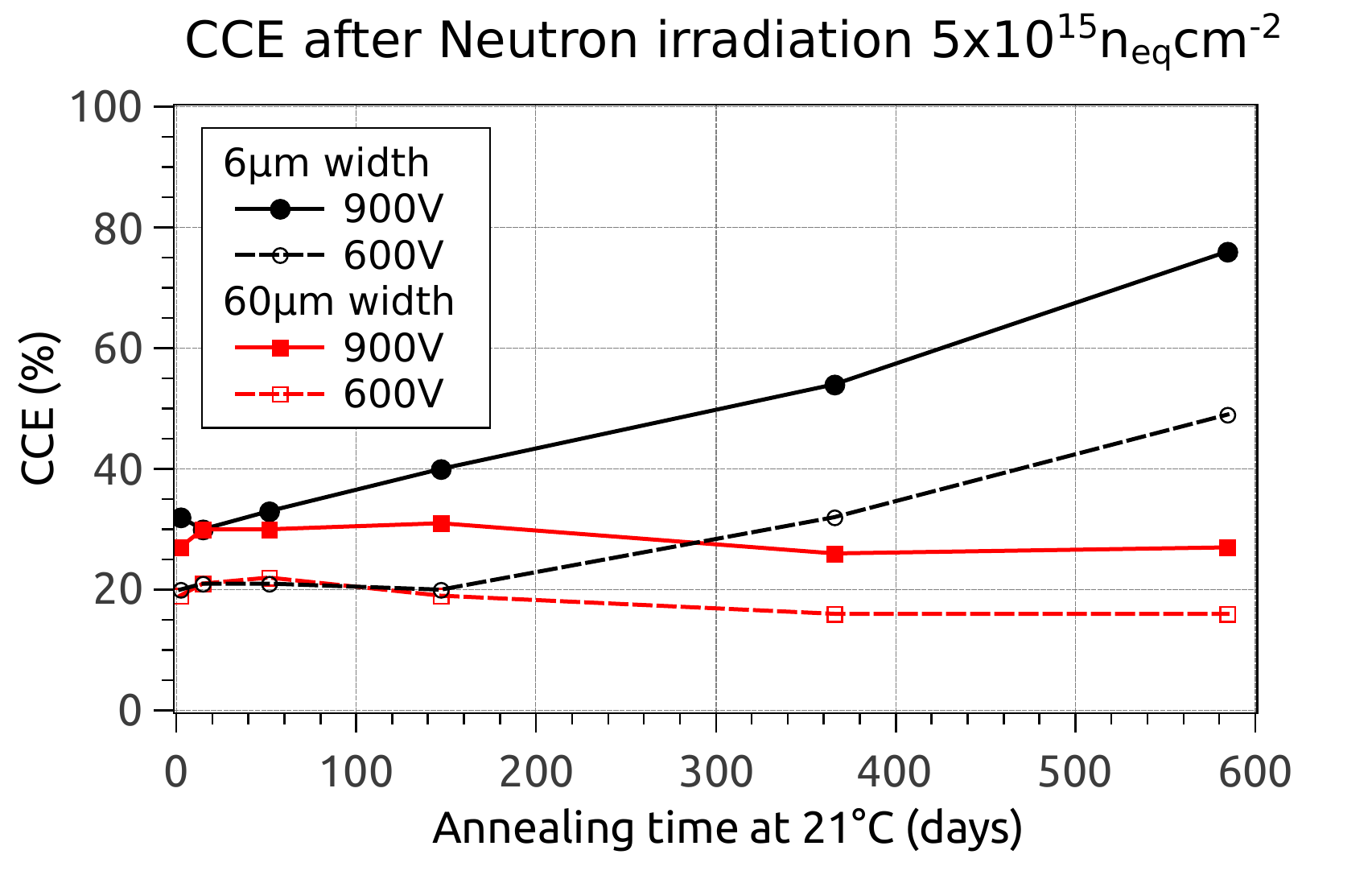}%
\caption{Charge collection efficiency vs. equivalent annealing time at room temperature for p-in-n strip sensors with p-spray strip isolation exposed to 23~MeV protons (left) and reactor neutrons (right)~\cite{Dierlamm}. Data are shown for different w/p ratios. The pitch amounts to 80\,$\mu$m, while the width is denoted in the figure legend.}%
\label{fig:CM_CCE}%
\end{figure}

This statement might hold until special structures of special designs enhance the electric fields at the collecting electrodes, thus generating charge multiplication in certain constellations. Within the CERN RD50 Collaboration both working groups contributed to a study to understand the relation of charge multiplication to sensor design and irradiation. The outcome was that a width over pitch ratio, w/p, of $\leq$ 0.15 is required to see increased charge collection at high voltages after neutron irradiation and long annealing (Fig.~\ref{fig:CM_CCE}). Additional induced oxide charges from charged particles suppress the electric fields at the surface and therefore prevent charge multiplication.

\begin{figure}%
\centering
\includegraphics[width=0.6\columnwidth]{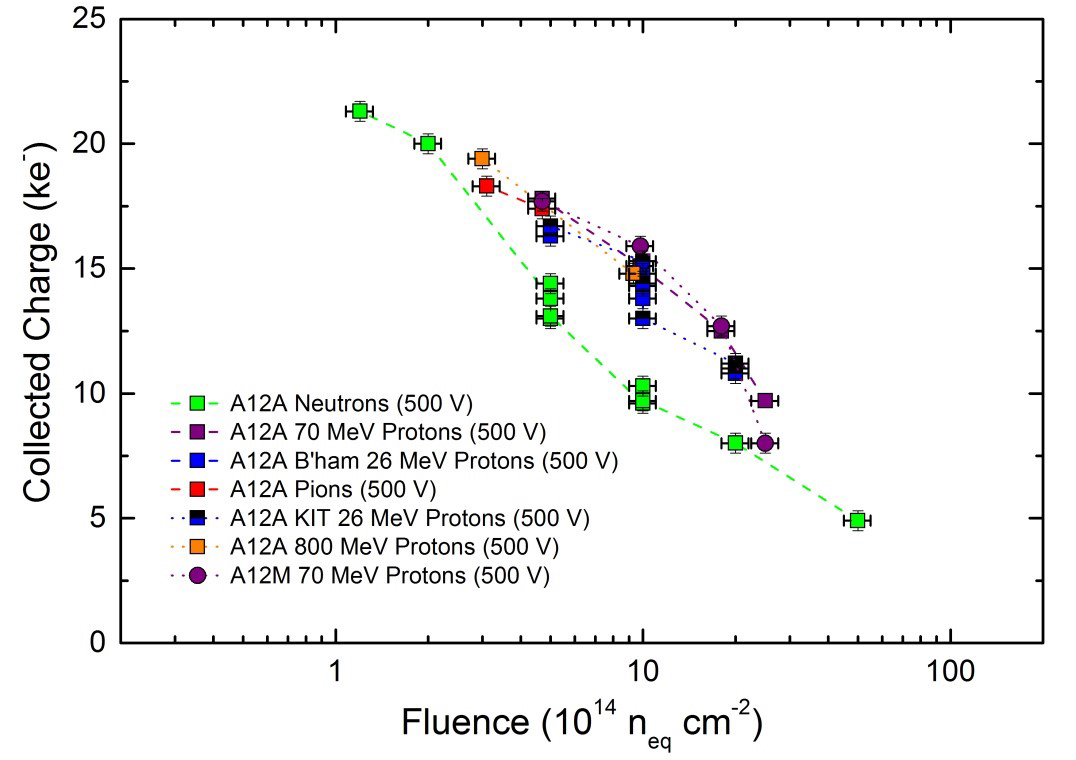}%
\caption{Collected cluster charge vs. fluence generated by several particle types. The measurements have been performed on strip sensors from HPK produced for ATLAS. The FZ material is expected to be oxygenated~\cite{Wonsak}.}%
\label{fig:CC_particles}%
\end{figure}

Both working groups did also observe that n-in-p type FZ strip sensors show superior charge collection at moderate bias voltages ($\sim500$\,V) after charged hadron irradiation compared to neutron irradiation (Fig.~\ref{fig:CC_particles}). This seems to be correlated with higher oxygen concentration in the investigated sensors compared to previously investigated samples (e.g. by RD50).
HPK sensors produced from \textquotedblleft standard\textquotedblright FZ material obtained by the CMS collaboration have been characterised by means of Secondary Ion Mass Spectroscopy (SIMS). The oxygen concentration was measured to be $9\times 10^{16}$~cm$^{-3}$. This is nearly an order of magnitude higher than expected from previous sensor production runs. Also ATLAS received similar \textquotedblleft standard\textquotedblright FZ material from HPK. It is well known that oxygenated material shows a lower full depletion voltage after charged particle irradiation compared to neutron irradiation. The high oxygen content in \textquotedblleft standard\textquotedblright FZ silicon has not been observed before.

\begin{figure}
\centering
\includegraphics[width=0.7\columnwidth]{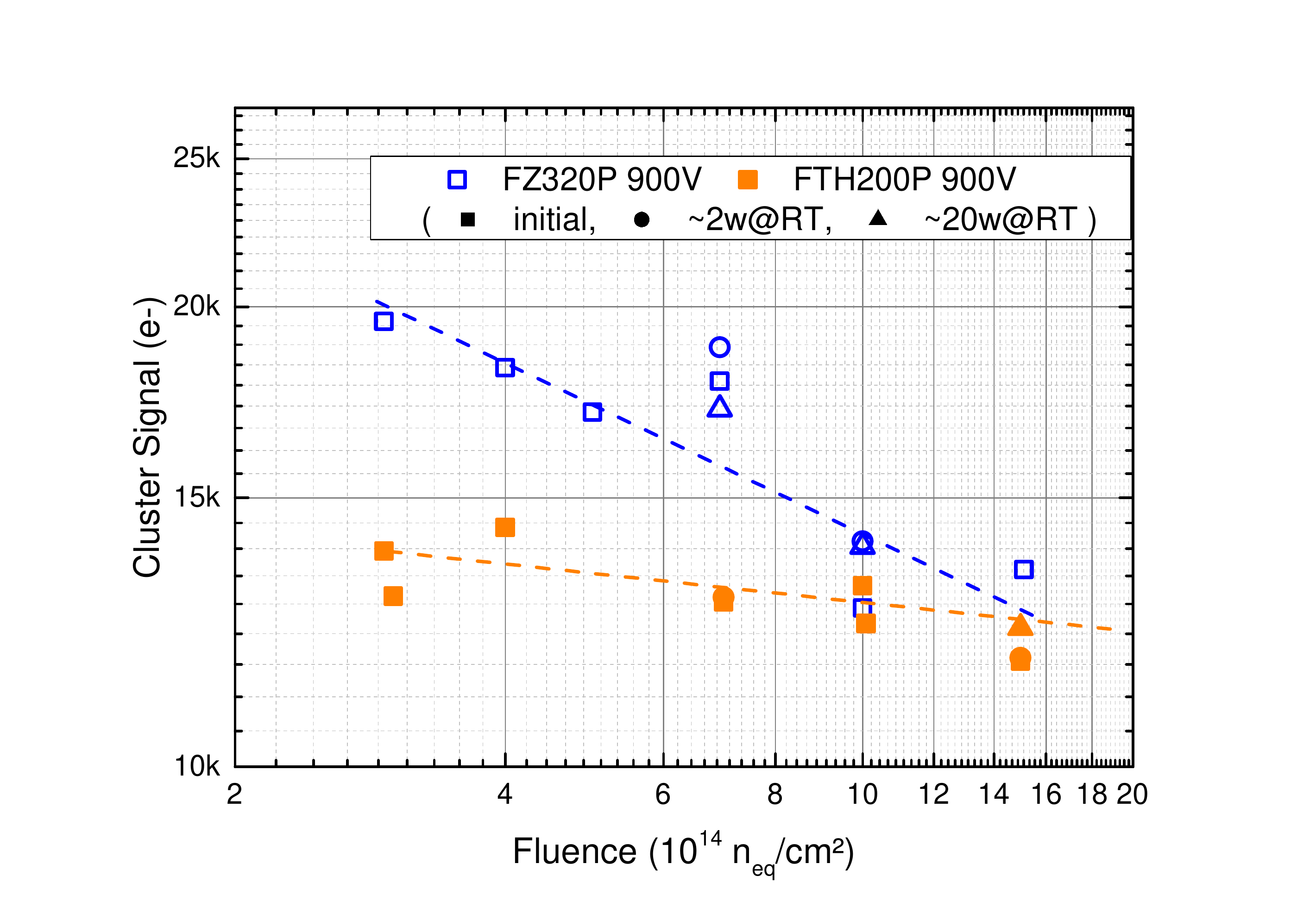}
\caption{Cluster signal versus fluence for p-type float-zone sensors operated at a bias voltage of 900\,V. Sensors with 320\,$\micro\meter$ thickness are shown in blue, while sensors with 200\,$\micro\meter$ thickness are shown in orange. The initial values measured right after irradiation are shown as squares, while results of measurements after two weeks and 20 weeks annealing at room temperature (RT) are shown as circles and triangles, respectively.}
\label{fig:CC_thickness_1}
\end{figure}

\begin{figure}
\centering
\includegraphics[width=0.6\columnwidth]{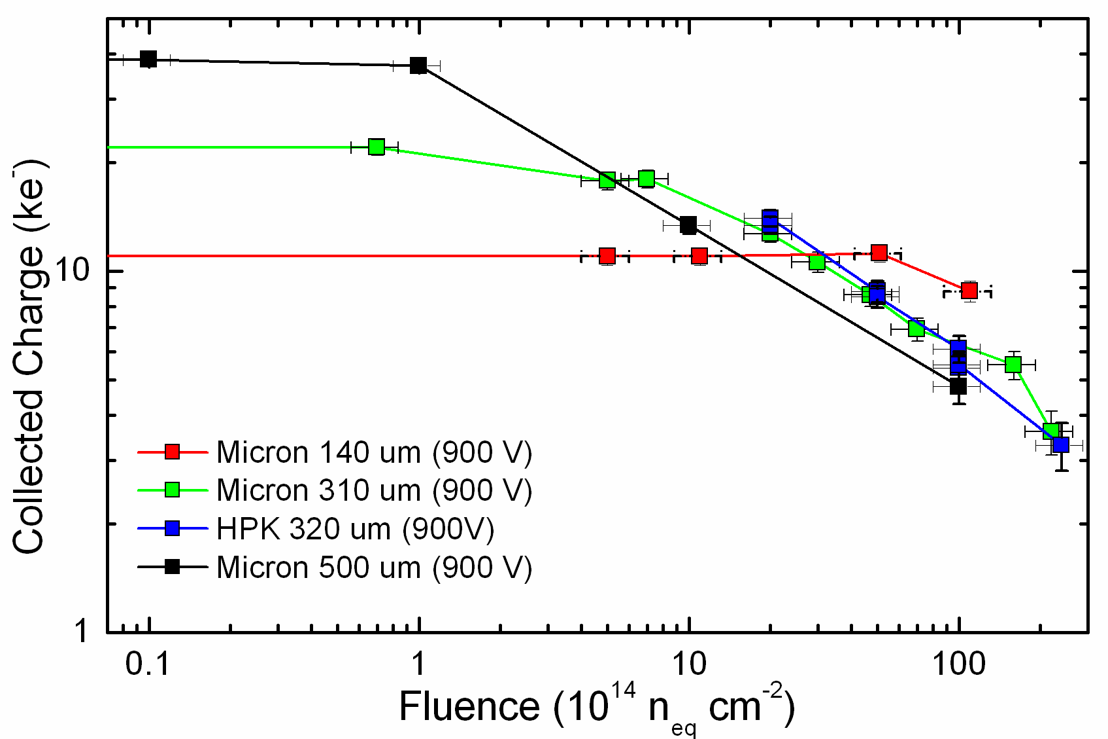}%
\caption{Collected cluster charge versus fluence for silicon strip sensors of different thicknesses~\cite{Affolder}.}%
\label{fig:CC_thickness_2}%
\end{figure}

Another aspect of the silicon material is its thickness, which was again studied by both working groups. The measurements confirm that the charge collection of thicker sensors is higher at low fluences. Further measurements at higher fluences showed that the charge collection of thinner sensors becomes equivalent and even superior to that of thick sensors. This crossing of the collected charge happens for 320\,$\micro$m and 200\,$\micro$m thick sensors at $1-2\times 10^{15}$~n$_{\mathrm{eq}}$\,cm$^{-2}$ (Fig.~\ref{fig:CC_thickness_1}) and for 320\,$\micro$m and 140\,$\micro$m thick sensors at $3-5\times 10^{15}$~n$_{\mathrm{eq}}$\,cm$^{-2}$ (Fig.~\ref{fig:CC_thickness_2}).

\subsection{Defect models}
\label{sec:models}
An effective two-trap model for simulations with Synopsys Sentaurus has been developed in order to describe radiation damage in silicon sensors. Since NIEL-violation was found for neutron and proton irradiation, the model has to introduce two different sets of parameters for both types of irradiation. The model aims to simulate the leakage current and depletion voltage and to provide information about the electric field. The model is not optimized for fluences above $1\times 10^{15}$\,n$_{eq}$\,cm$^{-2}$ and requires further tuning to match the corresponding expected sensor degradation. A valid radiation model is not only important for the simulation of the response of an irradiated sensor during the R\&D phase of a future HL-LHC detector, but also for the quantitative description of the signal loss in irradiated silicon detectors for Monte-Carlo simulations. 

Starting from a two trap model that was originally developed for the description of the electric field in sensors after proton irradiations~\cite{Ere04}, the model has been tuned to describe neutron and proton damage in silicon in the fluence range from \SIunits{$1\times$\power{10}{14}}{\,\cm \rpsquared} to \SIunits{$1\times$\power{10}{15}}{\,\cm \rpsquared}. The model uses two effective traps, a donor and an acceptor level. The trap parameters are the energy, the defect concentration as a function of the fluence and the electron and hole cross sections ($\sigma_{e,h}$). The parameters were determined by multi-parameter fits to capacitance-voltage (CV), current-voltage (IV) and TCT measurements performed on diodes within the framework of the CMS-HPK campaign. An overview of the parameters for neutron irradiation is presented in Table \ref{tab:01}, while the parameters for proton irradiation are shown in Table \ref{tab:02}. The complete work can be found in~\cite{ebe14}.

\begin{figure}
		\centering
		\includegraphics[width=0.45\linewidth]{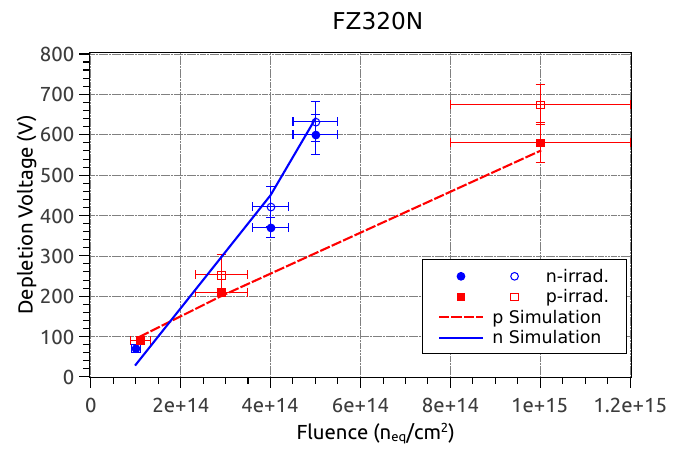}%
		\includegraphics[width=0.45\linewidth]{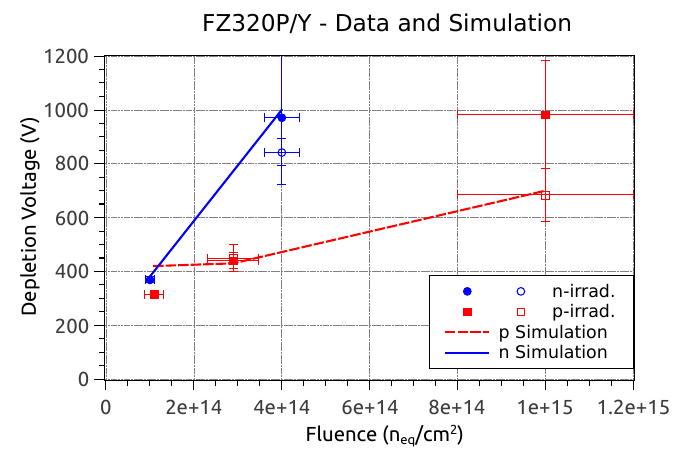}%
		\caption{Comparison of the depletion voltage extracted from CV measurements (markers) and from simulations (lines) for 320\,$\micro\meter$  float-zone n-bulk diodes (left) and p-bulk diodes (right) \cite{ebe14}.}
		\label{fig:sim1}
\end{figure}

The results from CV measurements of float-zone pad-diodes with 320\,$\micro\meter$ bulk thickness performed at $T=-20$ \SIunits{\degree}C at a frequency of 1\,kHz were compared to the simulated depletion voltages. Results are presented in Fig.\,\ref{fig:sim1} for n-type bulk (left) and p-type bulk (right). \\%
The model has been tuned for the sensor thicknesses of 320\,$\micro\meter$ and reactor neutrons or 23\,MeV protons, respectively, thus the agreement to measurements of thinner sensors or other irradiations is expected to be less accurate. However, for the sensors and irradiations under which the study was performed, the depletion voltages agree well. Moreover, the charge collection efficiency can be simulated with the effective two-trap model with an accuracy of 20\,\%.

\begin{table}[!t]
	\renewcommand{\arraystretch}{1.3}
	\caption{Effective two-trap model for neutron irradiation \cite{ebe14}. $E_V$ and $E_C$ denote the upper and lower edges of the valence and conduction bands, respectively, while $\Phi_{eq}$ denotes the equivalent fluence.}
	\label{tab:01}
	\centering
	\begin{tabular}{|c|c|c|}
		\hline
		Parameter & Donor & Acceptor \\%
		\hline
		Energy [eV] & $E_{V}+0.48$ & $E_{C}-0.525$ \\%
		Concentration [\SIunits{\cm \rpcubed}] & 1.395 \SIunits{\reciprocal \cm}$ \times \Phi_{eq}$ &1.55 cm$^{-1} \times \Phi_{eq}$  \\%
		$\sigma_{e}$ [\SIunits{\cm \squared}] & $1.2\times$\power{10}{-14} & $1.2\times$\power{10}{-14}\\%
		$\sigma_{h}$ [\SIunits{\cm \squared}] & $1.2\times$\power{10}{-14} & $1.2\times$\power{10}{-14}\\%
		\hline
	\end{tabular}
\end{table}%

\begin{table}[!t]
	\renewcommand{\arraystretch}{1.3}
	\caption{Effective two-trap model for proton irradiation \cite{ebe14}. $E_V$ and $E_C$ denote the upper and lower edges of the valence and conduction bands, respectively, while $\Phi_{eq}$ denotes the equivalent fluence.}
	\label{tab:02}
	\centering
	\begin{tabular}{|c|c|c|}
		\hline
		Parameter & Donor & Acceptor \\%
		\hline
		Energy [eV] & $E_{V}+0.48$ & $E_{C}-0.525$ \\%
		Concentration [\SIunits{\cm \rpcubed}] & 5.598 \SIunits{\reciprocal \cm} $\times \Phi_{eq}$ $-3.949\cdot$\power{10}{14}& 1.189 \SIunits{\reciprocal \cm} $ \times \Phi_{eq}$ + $6.454\cdot$\power{10}{13}  \\%
		$\sigma_{e}$ [\SIunits{\cm \squared}] & $1.0\times$\power{10}{-14} & $1.0\times$\power{10}{-14}\\%
		$\sigma_{h}$ [\SIunits{\cm \squared}] & $1.0\times$\power{10}{-14} & $1.0\times$\power{10}{-14}\\%
		\hline
	\end{tabular}
\end{table}%

\subsection[Bulk defects]{Investigations on bulk defects}
\label{sec:defects}
It is assumed that the more accurate the defect models are, the better the simulations will agree with measurements. And although an amazing progress has been achieved in the quality of the simulations over the recent years, there are still some open questions especially regarding the generation of space charge in high-energy proton-irradiated sensors. A forecast of detector properties can be achieved by investigating the defects in the crystal bulk.

Previous studies (\cite{pin08}-\cite{rox13}) have identified several defects with an impact on the detector properties. A NIEL-violation for proton-irradiated oxygen rich sensors was found. The reason for this was suspected in a higher donor generation in oxygen-rich silicon. For irradiation with different proton energies a similar effect on the depletion voltage was observed \cite{erf14}.

\begin{figure}%
		\centering
		\includegraphics[width=0.8\linewidth]{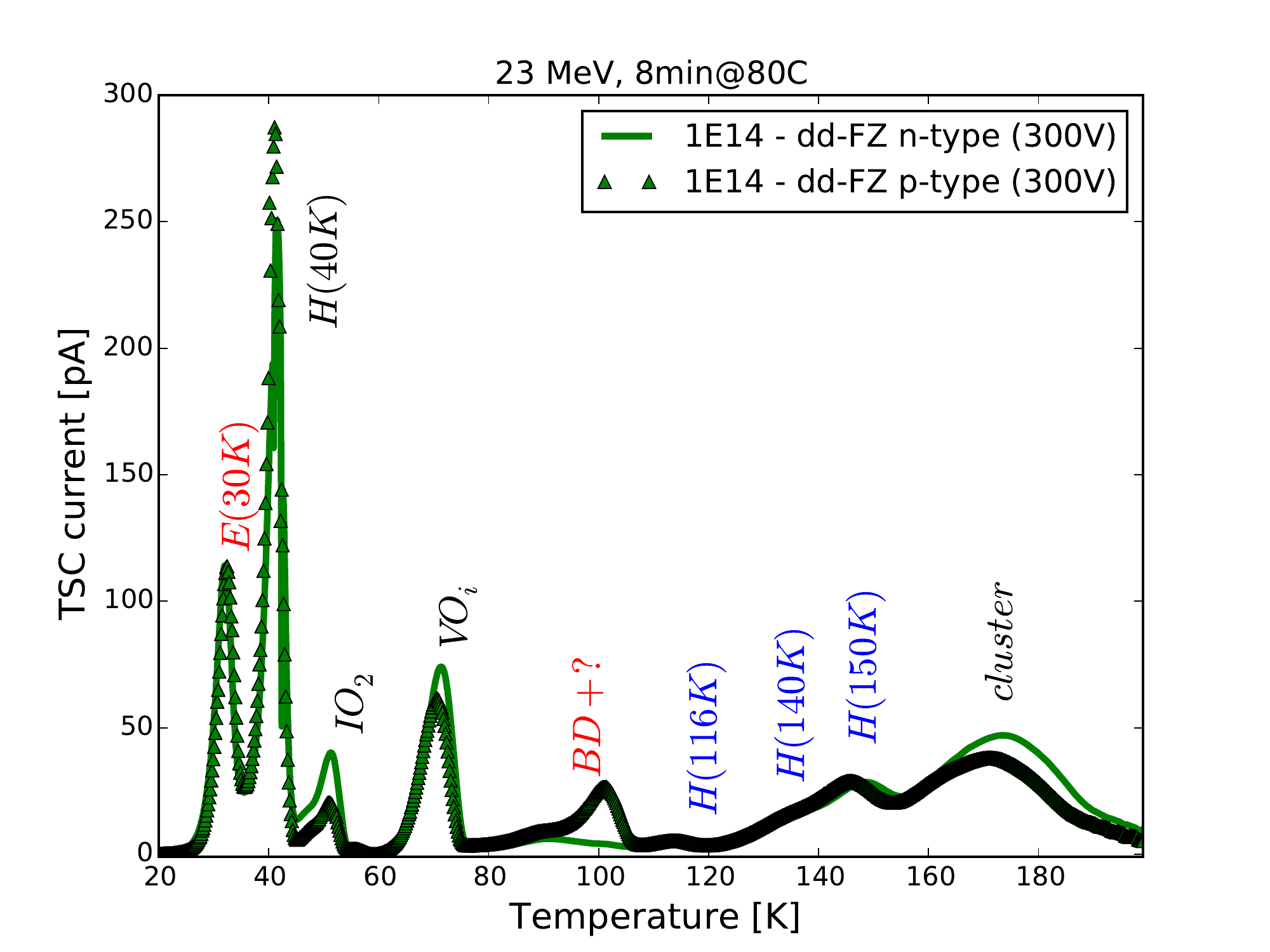}%
		\caption{ TSC spectra for 23 MeV proton irradiated n- and p-type dd-FZ pad-diodes with a fluence of $1\times 10^{14}$\,n$_{eq}$\,cm$^{-2}$.}%
		\label{fig:def1}
\end{figure}

Previous measurements were only performed on n-bulk devices and the understanding of the damage was therefore also based purely on n-type silicon. Recently, \textit{thermally stimulated current} (TSC) technique measurements on n- and p-bulk pad-diodes confirmed the educated guess that the defect generation in n- and p-bulk sensors is very similar. Figure \ref{fig:def1} presents the comparison of two TSC spectra for 23\,MeV proton irradiated n- and p-type deep-diffused float-zone (dd-FZ) pad-diodes with an oxygen concentration of [O]\,=\,\SIunits{$7\times$\power{10}{16}}{ \cm \rpcubed} and a fluence of $1\times 10^{14}$\,n$_{eq}$\,cm$^{-2}$.

Defects which are charged at room temperature contribute to the change of the space charge. These defects are marked with colored labels in Fig.\,\ref{fig:def1}: two donor levels (E(30K) and BD-defect) as well as a group of deep acceptors (H(116K), H(140K) and H(150K)) were observed for n- and p-type sensors\footnote{ The additional impact of the leakage current on the space charge and resulting double junction effect is not addressed here.}. In p-type silicon, overlapping with the BD-defect, an un-identified defect is seen at around 100\,K. This is likely the interstitial boron-oxygen complex ($B_{i}O_{i}$), a defect related to the doping of the bulk material. It has been reported \cite{tro80, kim91} to be a donor at E$_{C}-0.23$\,eV. The so-called acceptor removal might be related to the appearance of this defect; in turn this might help to understand the difference between the initial radiation damage as a function of the fluence in n- and p-type material. In order to confirm this assumption, more tests are required.

\newpage

\section{Low Mass System Design}

With their new tracking detectors both ATLAS and CMS will reduce the amount of passive material while at the same time maintain an excellent tracking performance and adding functionality to the front-end. For the high-luminosity era of the LHC ATLAS will build an all-silicon tracker consisting of several strip detector layers. CMS will utilize so-called $p_T$-modules that allow for a $p_T$ discrimination on the module for data reduction and the use of tracker information in the level 1 trigger of the experiment. The future trackers of both experiments will consist of a barrel section and two end-caps each.

\begin{figure}%
		\centering
		\includegraphics[width=0.9\linewidth]{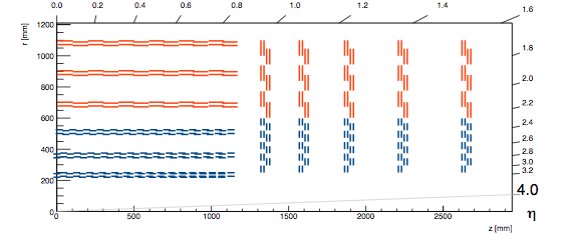}%
		\caption{Layout of the future CMS Tracker. A cut through one quarter of the detector in the $r-z$ plane is shown. The blue and red lines indicate PS and 2S modules, respectively.}%
		\label{fig:WP3a}
\end{figure}

\begin{figure}%
		\centering
		\includegraphics[width=0.75\linewidth]{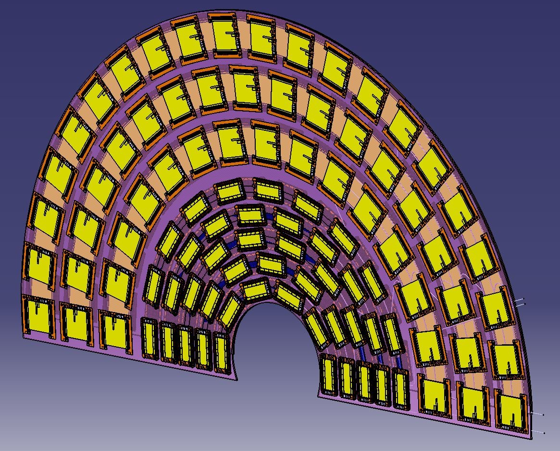}%
		\caption{CAD design view of a CMS end-cap half disk (\textquotedblleft dee\textquotedblright) equipped with modules.}%
		\label{fig:WP3b}
\end{figure}

The current baseline layout for CMS is shown in Fig.\,\ref{fig:WP3a}. Above a radius of 60\,cm the tracker is equipped with strip-strip modules (2S, indicated in red) each with an active area of approximately 10\,cm x 10\,cm. They consist of two silicon strip sensors arranged on top of each other. The signals from both sensors are read out by a common front-end ASIC that correlates the signals and sends out trigger information for tracks with $p_T$ above a programmable trigger threshold. In the inner radii down to 20\,cm the tracker will be equipped with pixel-strip modules (PS, indicated in blue) consisting of a strip and a pixel sensor. The PS modules provide hit position information also in the third coordinate ($z$) and are capable of $p_T$ discrimination as well. The 2S modules have five dedicated cooling contacts which are used to mount the modules on their support structure and efficiently attach them to the cooling circuitry. The PS modules are directly mounted to their support structure via a phase transition thermal interface material that allows for a large-area and low impedance thermal contact. The end-caps of the future CMS tracker will consist of large half-disks with a diameter of 1.2\,m - so-called dees - that are arranged in double disks. The dees are equipped with modules on both sides. Figure\,\ref{fig:WP3b} shows the CAD design view of a dee. Two dees form a disk and two disks in turn form a double-disk. The overlap between modules in $\phi$ is established by alternating the module $\phi$ positions on the front and back side of a dee, whereas the overlap in radius is ensured by alternating the radial positions of the modules on the two disks within a double-disk. Each end-cap will consist of five double-disks.

\begin{figure}
		\centering
		\includegraphics[width=0.78\linewidth]{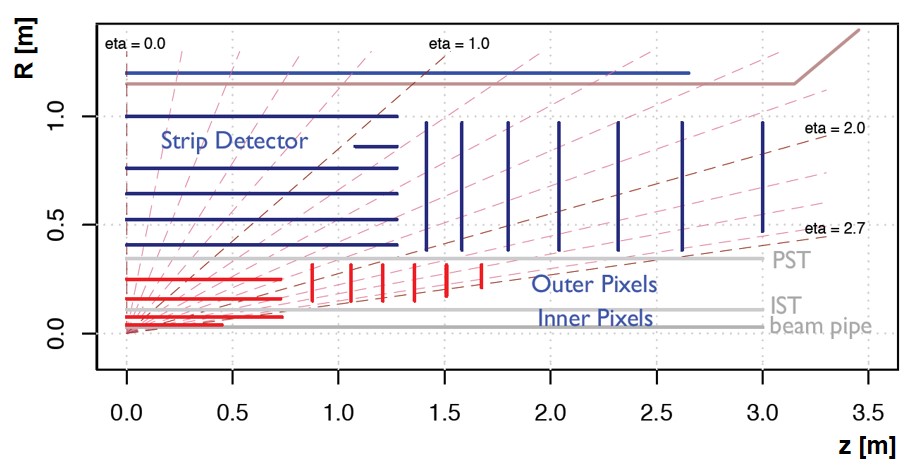}
		\caption{Layout of the future ATLAS tracker. A cut through one quarter of the detector is shown in the $r-z$ plane. The strip tracker is shown in blue.}
		\label{fig:WP3c}
\end{figure}

The future ATLAS tracker will have a very similar layout compared to that of CMS. The layout is shown in Fig.\,\ref{fig:WP3c}. The pixel detector will extend to radii up to about 35\,cm. At larger radii silicon-strip detector modules will be used throughout the tracker. All module types will be directly glued to their support structure. In the barrel the support structures will consist of so-called staves arranged around the beam axis. The end-cap disks will be equipped with wedge-shaped petals of a size of approximately 40\,cm x 60\,cm. Details on the stave and petal design can be found in Sect.\,\ref{sect:atlasrequirements}.

The ATLAS support structures - staves and petals - as well as the CMS barrel support structures for PS modules and the end-cap dees are highly integrated sandwich structures to which the detector modules are directly attached. The core of the structures consist of a light-weight carbon or polyimide honeycomb (ATLAS) or polyetherimid foam (CMS) for structural purposes, and carbon foam heat spreaders into which the cooling pipes are embedded. The facings of the sandwich are made of high-modulus carbon fibre with thermal conductivity of above 600 W/m/K. With the appropriate layup of the carbon composite the facings act as the dominant path for the heat produced by the modules while ensuring sufficient stiffness of the structure itself.

Due to the increased instantaneous luminosity and therefore higher track densities, both experiments will significantly increase the granularity in order to keep the hit occupancy at an acceptable level. This results inevitably in an increase in front-end power, which calls for novel powering schemes. Both ATLAS and CMS will use DC-DC power converters close to or directly on the detector modules in order to reduce the overall cable cross section for powering and lower the resistive losses in the cables.
As outlined in Section\,3, the silicon sensors have to be radiation hard in order to withstand the expected dose corresponding to 3000\,fb$^{-1}$ of integrated luminosity that will be collected during HL-LHC running. This requires operation of the sensors at temperatures around or below $-20^{\circ}$C to mitigate radiation damage effects. Following developments for the LHCb Velo detector, ATLAS and CMS will use two-phase CO$_2$ cooling for their entire tracking detectors. This will not only ensure efficient cooling but also reduce the amount of passive material due to smaller pipe diameters and wall thicknesses, and the larger radiation length, $X_0$, of the coolant itself.

Prototype work for ATLAS petals and staves as well as CMS end-cap structures have started and will confirm the mechanical and thermal design of the novel concepts.

\section{Automated Precision Assembly Procedures}

\subsection{Introduction}
In the work package ``Automated Precision Assembly Procedures'' both requirements and possible solutions for precise assembly of silicon detectors were investigated. 
Silicon detectors achieve micrometre precision in the measurement of particle tracks. Although software based alignment corrections have proven to be very powerful, the assembly precision of the detectors is still essential. When silicon tracking information is to be used at the first trigger level, software alignment is, depending on the implemented scheme, difficult or impossible to use. Moreover, available clearances have to be matched during installation. On the other hand, the new tracking systems will be composed of tens of thousands of silicon sensors. The development of automated and precise assembly techniques is required and can partly be based on the experience obtained in the construction of the current tracking systems. In the following the investigations within this work package are described. First the experience from the ATLAS and the CMS experiments has been collected and compared, and the requirements for the new detector systems have been analyzed. Secondly, ideas for precision assembly of silicon sensors into modules and larger structures are discussed and first tests on prototypes are layed out. Additionally, the possibility of assembly in industry is surveyed.
% A method for the precise metrology of double sided objects in which silicon sensors have to be aligned with high precision back-to-back is under development.

\subsection{Requirements}

\subsubsection{ATLAS experiment \label{sect:atlasrequirements}} 

For the upgrade silicon strip tracker of the ATLAS experiment the barrel region is planned to be built of three short strip layers with strip lengths of about 24\,mm, and two strip layers with longer strips being about 48\,mm long (Fig.\,\ref{fig:WP3c}). A stub cylinder is  foreseen in the transition region between the barrel and forward tracker regions.
The end-caps, in the forward region, are consisting of seven strip layers with varying strip length of 8-48\,mm. Many details on the layout are given in reference~\cite{loi}.
In total, the strip detector will consist of 191\,m$^2$ of silicon with about 74 million channels. This results in the need of about 20\,000 modules. Currently, there are no specific trigger layers planned. But the tracking information is used for triggering by selecting regions-of-interest which are built from muon and calorimeter information by the level-0 trigger and fed into front-end chips of the silicon tracking detector.  
The silicon strip system will  have a size of about six metres in length and 1.1\,m in diameter. These parameters are given by the existing ATLAS detector and result in small clearances of about 1.5\,cm between the calorimeter and the new strip detector system. It sets already the first requirement of precision - the new detector has to fit in. 
% requirement document
%
%\begin{figure}
%\begin{center}
%\includegraphics[width=.6\textwidth]{loil.pdf}
%\caption{The baseline layout of the replacement tracker showing the active areas of silicon detectors arranged on cylinders and disks.}
%\label{loil}
%\end{center}
%\end{figure}
The main concept in the upgrade ATLAS strip tracker is modularity. Meaning, components are built into stand-alone functional objects, and those are put together in a second step. This aims for easier final assembly and multiple site production.
%In the following this will be explained exemplary for the end-cap region. 

\begin{figure}
\begin{center}
 \includegraphics[width=.5\textwidth]{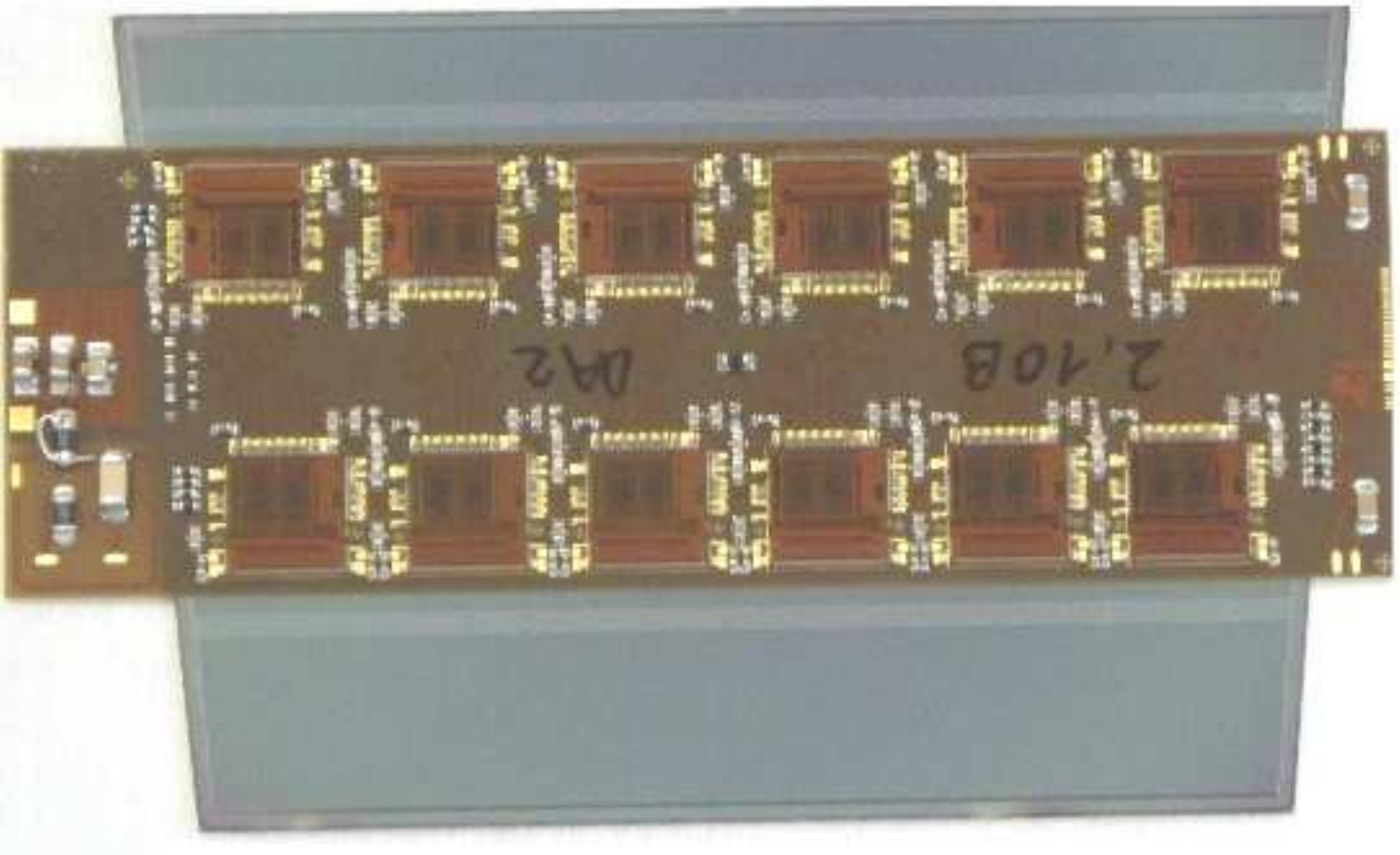}
\caption{Photo of a prototype end-cap module consisting of a silicon senor and a flex kapton hybrid on top~\cite{kuehns}.}
\label{bmodule}
\end{center}
\end{figure}

\begin{figure}
\begin{center}
\includegraphics[width=.8\textwidth]{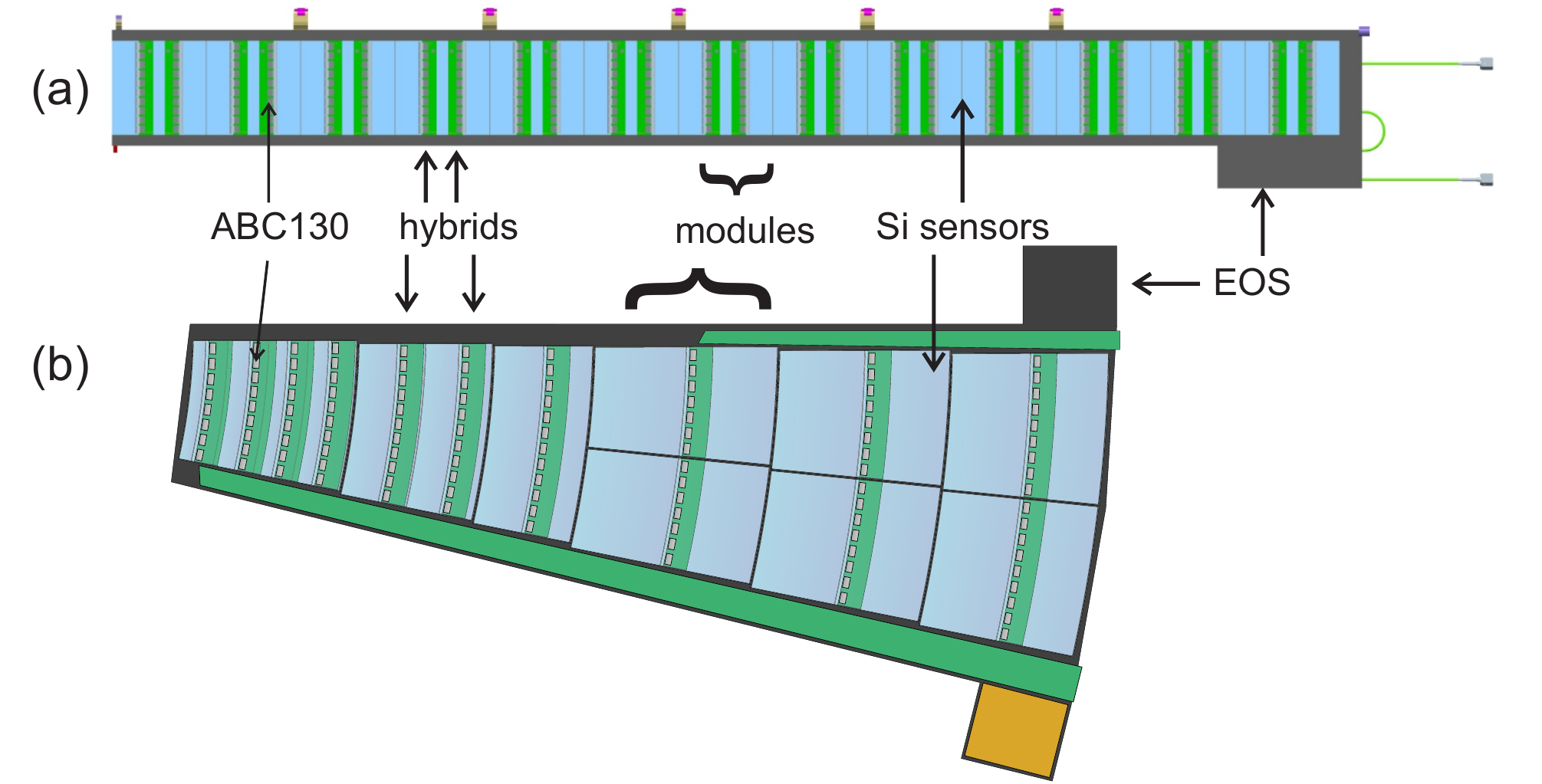}
\caption{(a) Barrel stave components, (b) petal components~\cite{loi}. The abbreviation \textquotedblleft EOS\textquotedblright refers to the end-of-stave (or more general end-of-structure) region.}
\label{strips}
\end{center}
\end{figure}

The basic principle is that modules are assembled from a silicon sensor with a hybrid glued on top. A photograph of one prototype module for the end-cap region is shown in Figure~\ref{bmodule}~\cite{kuehns}. It shows a silicon sensor with a kapton flex hybrid glued on top, which reduces the amount of support material. The displayed module uses 12 ASICs produced in a 250\,nm CMOS process, which are glued onto and wire-bonded to the hybrid. For the barrel region similar modules are in prototyping with rectangular sensor shape.
In the second step, modules are glued on support structures made of carbon and honeycomb as can be seen in Figure~\ref{strips}~\cite{loi}.

\begin{figure}
\begin{center}
\includegraphics[width=.75\textwidth]{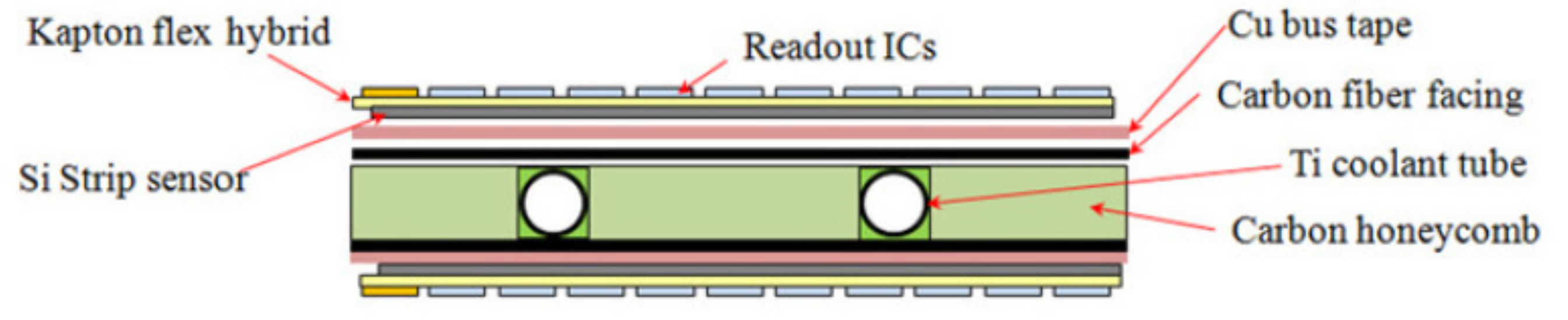}
\caption{Cross section of a stave~\cite{sergio}.}
\label{stavecross}
\end{center}
\end{figure}

\begin{figure}
\begin{center}
\includegraphics[width=.65\textwidth]{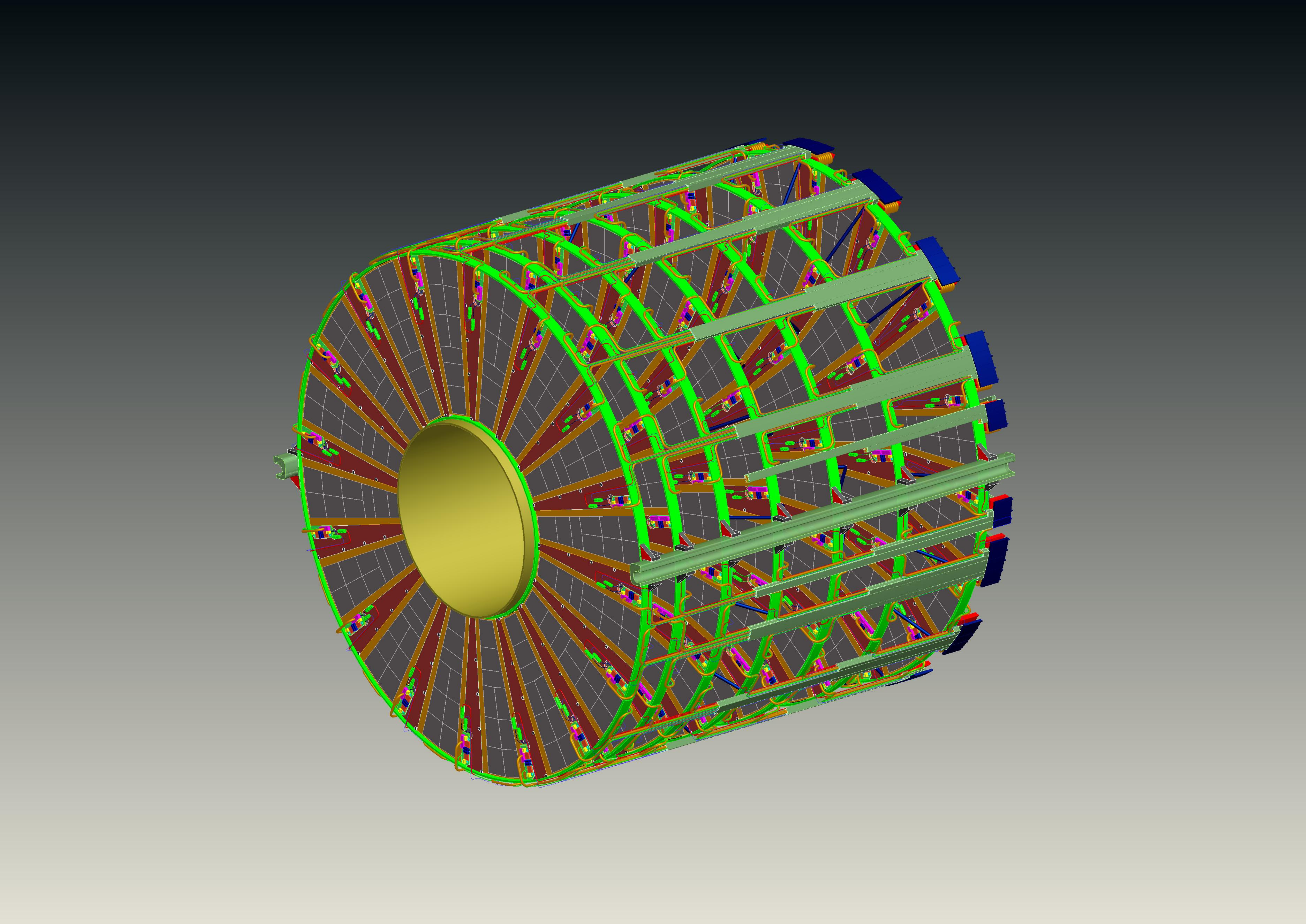}
\caption{Drawing of one end-cap showing the carbon support structure (light green), seven disks with 32 petals (grey) each and service modules (dark green).}
\label{endcap}
\end{center}
\end{figure}

In part (a) of this figure a so-called stave for the barrel region is shown. It is a structure of 1.3\,m length with modules assembled onto both sides. Part (b) sketches the analogue for the end-cap region, a so-called petal. It has a wedge-shape and a length of about 0.6\,m. Routing is provided by kapton tape with copper traces below the modules. A cross-section is given in Figure~\ref{stavecross}~\cite{sergio}.
The single-sided modules are assembled onto the top and bottom of the support structure with a stereo angle of 40\,mrad. Both petals and staves are foreseen to be mounted in carbon support structures resulting in individual end-caps and a barrel. A drawing of one end-cap is shown in Figure~\ref{endcap}~\cite{kuehns}. It is planned that the whole system will be cooled with CO$_{2}$ to about $-20$\,$^{\circ}$C.

The physics and performance aims set several requirements on this modularly assembled detector system.
For detailed determination of requirements, an ATLAS internal working group for alignment requirements has been set up in Spring 2014. This group is aiming for a document for assembly which will give many details.
The following considerations result from private discussions within the before mentioned group and have to be taken with caution. 
To maintain performance, precise assembly of modules (silicon sensors) is needed on the carbon cores.
The baseline is 100\,$\micro\meter$ precision between sensors on adjacent cores and also back-to-front on the same core. The value also gives the maximum difference between true positions and known positions after placement and surveys. Therefore individual module placement requires a higher precision. Important to note is that a very precise survey of sensor and core positions is essential since precise alignment is foreseen using software. Also the minimum overlap of silicon strips has to be maintained. Moreover, the stability of the components, namely modules on cores and cores in support structures and of the support structure itself is very important and has to be determined from measurements.
Therefore data will have to be collected at different temperatures and for all petals and staves. This will allow temperature scaling of survey data.

From experience of the current tracker, it is assumed that a very detailed survey is preferred to a high-precision assembly in the ten micron-range. Therefore the survey is required to have precision comparable to detector resolution.

\subsubsection{CMS experiment}
With the high radiation damage in the sensors of the current strip modules and the harsh pile-up conditions at the HL-LHC the need for a new Tracker for the CMS Experiment is evident. The layout of the Tracker and especially the design of the modules are driven by the idea to include the information of reconstructed tracks already at the first triggering stage (L1) of the CMS experiment. The combination of the traditional calorimeter and muon system based triggers with the tracking information will result for example in better lepton isolation determination. Therefore low lepton $p_T$-thresholds comparable to current values are achievable that would otherwise not be possible in scenarios with 140 or more simultaneous proton-proton interactions in a single bunch crossing.

To allow the combination of track information with other trigger objects at L1, the reconstructed tracks have to be available at L1 within $\approx 5\,\micro\mbox{s}$ after the collision. The conventional offline track-fitting model used by CMS so far is not suited to achieve this. Additionally it is impossible to send the full hit information out of the tracking volume within that time, due to the limited bandwidth available. Therefore the hits that will be handed over to a hardware/FPGA based track reconstruction stage will be filtered in the front-end of the modules itself.

The criterion for the selection of hits is the transverse momentum. An estimate of the $p_T$ can be obtained by combining the hit information in two sensors located at a short distance ($2 - 6\,\mbox{mm}$). A vectorial property referred to as \textquotedblleft stub\textquotedblleft can be calculated, resembling the direction of the track at this space-point. Under the assumption that the origin of the track is the interaction point, one can reconstruct the circular motion of the particle and give a lower limit on the transverse momentum of the track. Presently stubs above a $p_T$ of $\approx 2 \,\mbox{GeV}$ are foreseen to be used in the trigger.

This concept drives the module design of the future CMS tracker. The modules will be equipped with two (in-plane) adjacent sensors read out by the same front-end electronics. This allows for fast spacial correlation of the binary hit information in the readout chips. Therefore the module itself will be able to select tracks for the trigger based on their transverse momentum and provide tracks above a certain threshold to the L1 trigger. If an event was accepted by the L1 trigger, the module will provide the full hit information to the back-end, including hits from low $p_T$ tracks.

The layout of the future tracker is shown in Fig.~\,\ref{fig:WP3a}. It comprises a classical barrel and end-cap geometry, featuring six barrel layers within a radius of about $25-110$\,cm and $2.5\,\mbox{m}$ in length. The five end-cap disks per side cover the same radial space and extend the cylindrical tracking volume to almost $6\,\mbox{m}$ in length. Each disk  hosts 15 rings of modules. The number of fundamentally different module types is kept to a minimum of two, with five geometrical variants in total. 

The inner layers of the barrel and the inner rings are instrumented with PS (pixel-strip) modules, featuring a pixel sensor and a strip sensor with $2.5\, \mbox{cm}$ long strips. The pixel size is $100\,\micro\meter \times 1.5\,\mbox{mm}$. These macro-pixels are introduced to provide a certain $z$-resolution.

\begin{figure}
\begin{center}
\includegraphics[width=.7\textwidth]{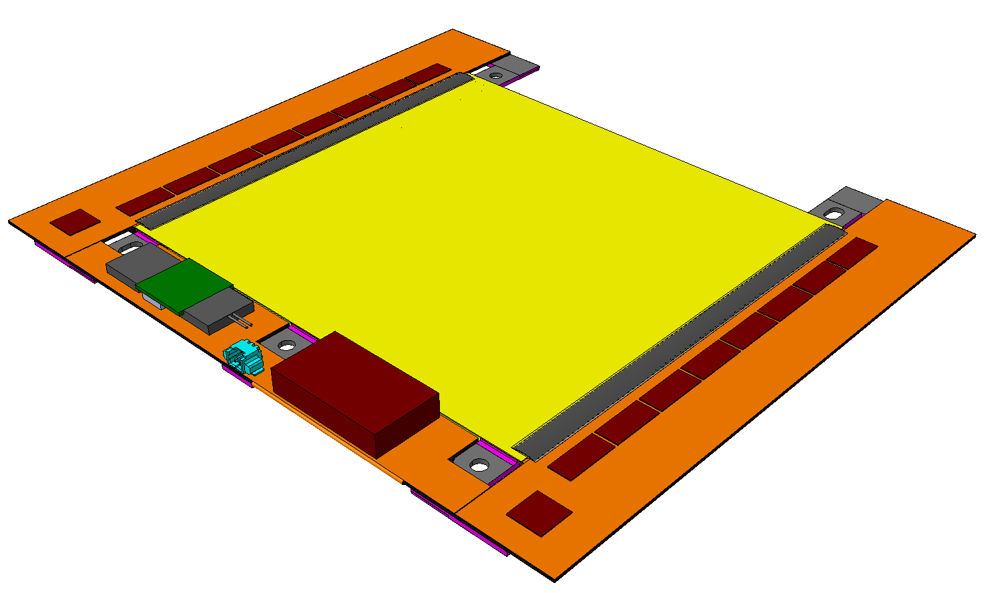}
\caption{CAD drawing of a 2S Module. The yellow area represents the sensors, the flex hybrids are shown in orange, the CBCs are shown as rows of eight red boxes, and the spacers are depicted in grey.}
\label{2SModule}
\end{center}
\end{figure}

In the outer regions about 8000 2S (two strip-sensor) modules are placed, featuring two identical sensors, each having two rows of 1016 strips of $\approx\,5\,\mbox{cm}$ length, resulting in a sensitive area of about $10\,\mbox{cm}\times\,10\,\mbox{cm}$. The sensors are placed onto Al-CF spacers that provide the mechanical support for the two readout hybrids and the service hybrid. The hybrids are attached to the sides of the sensor package, as shown in Fig.\,\ref{2SModule}. Eight CMS Binary Chips (CBC) are flip-chipped onto each readout hybrid, a flexible laminate that can be folded over the spacers, to allow for wire-bonding of both sensors to the same hybrid from top and below. Therefore one CBC must process the signals of 127 channels from the upper and 127 adjacent channels of the lower sensor, and correlate their signals to generate stubs.

Due to the success of the software track alignment, the requirements on the global module positioning inside the tracker by the offline track fitting are quite low and not even specified at this stage. A certain precision will be required by the tight clearances between modules, as needed to ensure high hermeticity, and is achievable by standard mechanical alignment procedures.
Within one module the required precision is much higher, since it is assembled in multiple steps and wire-bond connections are needed between most sub-assemblies, especially between the sensors and the readout hybrids.
The most sensitive constraint within the assembly of one module is the relative alignment between the strips of both sensors. Since within the CBC the offset and width of the region used to find the correlations can be programmed with half pitch resolution (i.e. $45\,\micro\meter$), lateral shifts between the sensors can be tolerated and corrected within the front-end itself. Rotational modes however cannot be corrected for, due to the fact that the strips are $5\,\mbox{cm}$ in length. Therefore it was decided to limit the angle between the strips of two sensors to $20\,\micro\meter$ over the $5\,\mbox{cm}$ length, resulting in an angle of $400\,\micro\mbox{rad}$. This results in an inaccuracy significantly smaller than the unavoidable inaccuracy in the stub generation caused by the limited spatial resolution of the sensor in combination with its planar nature.

During production the assembly precision has to be assured and monitored, which includes the necessity to compare the relative positioning of the upper and lower sensor. For this measurement alignment marks on both sensors need to be examined, ideally without having to turn over the whole module.

\subsection{Experience with module assembly for the current trackers}

\subsubsection{ATLAS experiment}

\begin{figure}
\begin{center}
\includegraphics[width=.7\textwidth]{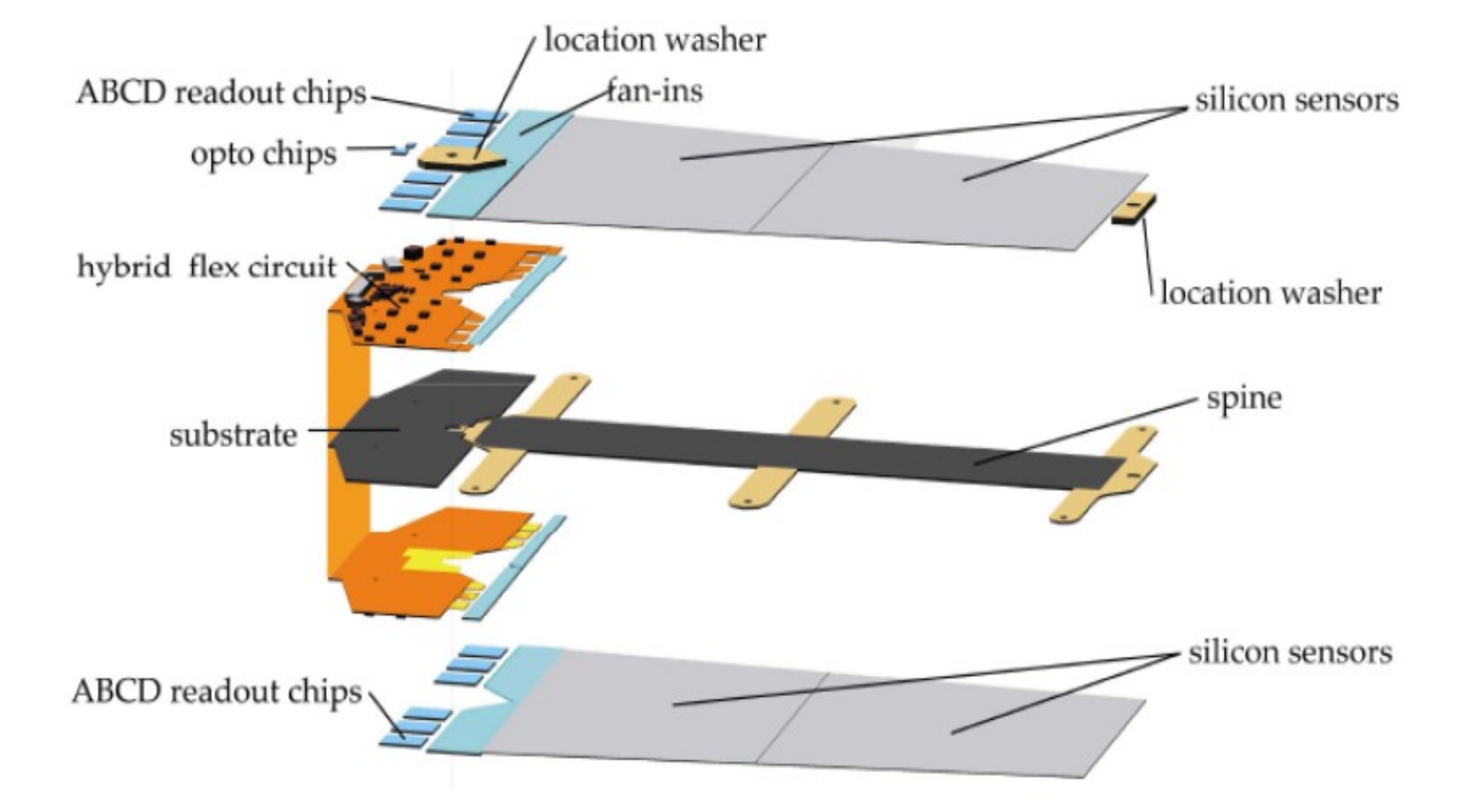}
\caption{Schematic drawing of a SCT end-cap module of the current ATLAS tracker~\cite{detpap}.}
\label{sctmod}
\end{center}
\end{figure}

The assembly of the current semiconductor tracker (SCT) of the ATLAS experiment was done by precise assembly of modules and mounting them onto disks at defined positions with screws. The modules consist of two silicon sensors glued on a thin, thermally conductive spine and flex hybrids wrapped around. A schematic drawing of a SCT end-cap module is shown in Figure~\ref{sctmod}~\cite{detpap}.

\begin{figure}
\begin{center}
\includegraphics[width=.6\textwidth]{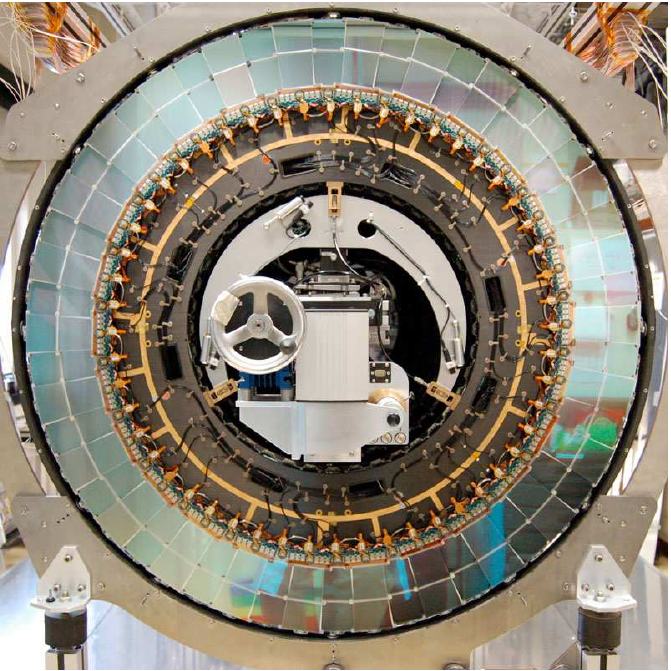}
\caption{Close-up photo of one disk of the SCT of the current ATLAS tracker~\cite{cern}.}
\label{sctdisk}
\end{center}
\end{figure}

The assembly of modules was performed manually at 14 different sites, which had to proof their ability before. Tested and well-performing modules were shipped to central sites, conducting module-to-disk assembly.
One disk assembled with modules is shown in Figure~\ref{sctdisk}~\cite{cern}.
Precision tools, optical alignment and fiducials were deployed to achieve a precise assembly of the modules. From the correlation obtained through fitting, a precision of end-cap modules of 17\,$\micro\meter$ in the $r-\phi$ coordinate and 580\,$\micro\meter$ in the $z$-coordinate was obtained.
Moreover, the mechanical tolerance for positioning sensors within the back-to-back pair was better than 8\,$\micro\meter$ transverse to the strip direction~\cite{atlindetec}. Barrel modules were partially assembled with a module mounting robot for module-to-stave assembly. Their mechanical back-to-back alignment   was less than 8 \,$\micro\meter$ (in-plane lateral, $x$), 20\,$\micro\meter$ (in-plane longitudinal, $y$), and 70\,$\micro\meter$ (out-of-plane, $z$, deviation from the average profile)~\cite{atlindetb}.
The series production lasted for two years, including 15\,\% spares. About 3\,\% of the total module production were out of $xy$ tolerance and about 2.6\,\% were out of $z$ tolerance. In total about 4600 modules were successfully built with yields of 90.5\,\% (barrel) and 93\,\% (end-cap). 

%- Alignement with offline track-based alignment uses a global chi-squared technique
%that minimizes the residuals to samples fitted tracks -> high accuracy but there
%are deformations of larger sectors, so called ``weak modes''.

\subsubsection{CMS experiment}
The present Outer Silicon Tracker of the CMS experiment consists of 15\,148 strip-sensor modules that are distributed in a cylindrical volume with a length of about $5.5$\,m and a diametre of $2.4$\,m. To ensure optimal coverage in the barrel and the forward regions, 29 module types were developed, using 15 different sensor designs and 12 readout hybrid variants. Despite this diversity of modules, they all share the same  characteristic layered build-up: a frame (carbon fibre or graphite) hosts a readout kapton circuit layer (FE hybrid) and one or two silicon strip sensors. In the latter case the two sensors are placed such that the strips are daisy-chained and the resulting sensor size exceeds the production 6 inch wafer size. The approximate size of a complete module varies between $6\times12~\mbox{cm}^2$ and $12\times23~\mbox{cm}^2$.

\begin{figure}
\begin{center}
\includegraphics[width=.6\textwidth]{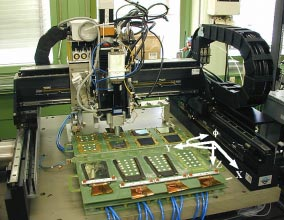}
\caption{Photograph of a gantry robot as used in the construction of silicon detector modules for the current CMS tracker~\cite{CMS_gantry}.}
\label{CMS_gantry_photo}
\end{center}
\end{figure}

Because of the common build-up of the modules the same assembly strategy was applicable to all module types, assembled at the worldwide distributed assembly centers. The assembly was done using automated assembly robots, so-called gantries~(Fig.~\ref{CMS_gantry_photo}). The same gantry type (Aerotech AGS 10000) was used as the main assembly utility at all production sites in order to achieve a uniform high throughput (eight modules per day) and a constant quality. This machine has a working area of about $50\times50~\mbox{cm}^2$ in plane, and a $z$-arm with a 10\,cm travel distance and a $360^{\circ}$ rotatable head. This arm could be equipped with glue dispensers and pick-and-place tools.

In the working area on a so-called base plate different platforms were placed, which had the purpose of holding the parts before and after assembly, utilizing a custom made embedded vacuum system.

Precision in the assembly process was gained by exploiting the high repeatability in the $xy$-plane in combination with a high resolution camera system. This was used to identify fiducial markers on all components before the pick-up, giving a reference between the picked up part and the assembly already finished in previous steps.

The actual assembly process was split into a manual and an automated part. The manual part included loading the module frames and the front-end hybrids onto the assembly plate and the equipping of the sensor platform with the sensors. Additionally  the glues had to be prepared and filled into the dispenser units of the gantry.
After the placement of the sensor and assembly platforms the gantry automatically performed the actual assembly of the parts. During a first run, the positions of all parts were measured, then glue was applied to the module frame, onto which the sensor and the hybrid were placed in subsequent steps. Before the assembly plate, carrying four modules and holding them in place by vacuum, was moved to an off-site storage for the long curing process, a first survey of the assembly was performed using the gantry to measure the achieved precision. After the curing process and the release of the fixation of the modules they were remeasured at an independent dedicated coordinate measuring machine.

\begin{figure}
\begin{center}
\includegraphics[width=.6\textwidth]{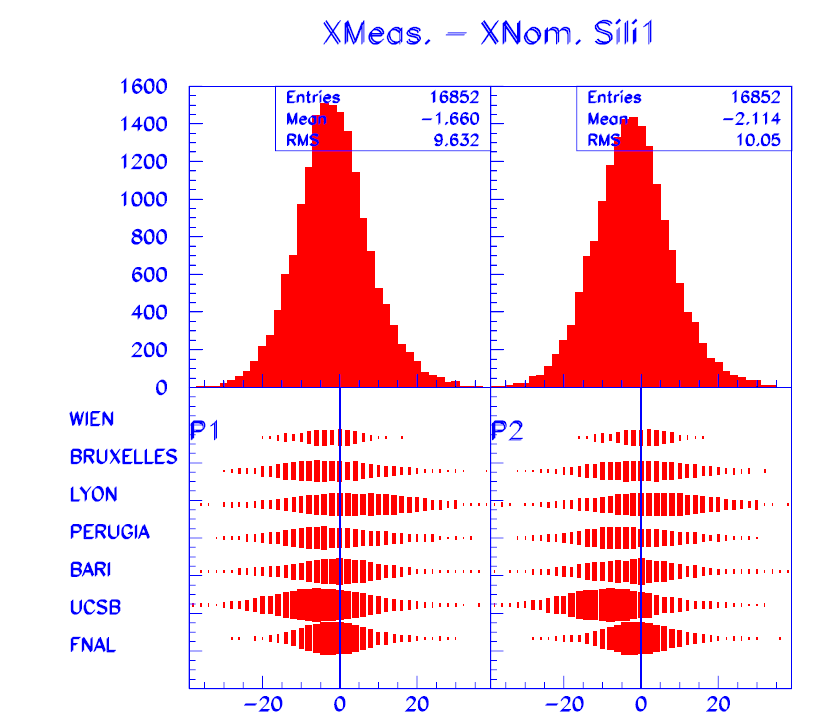}
\caption{Typical residual distributions of reference points on CMS strip modules, indicating a RMS of the order of $10\,\micro$m. The top distributions show the combined data from all gantry centers, while the lower plots show the achieved precision for the different cites individually \cite{CMS_experiment}.}
\label{CMS_gantry_precision}
\end{center}
\end{figure}

During both surveys the difference between the optimal sensor placement (deducted from markers on the frame) and the actual position of the sensor(s) in the $xy$-plane showed a distribution with a standard deviation smaller than $10~\micro\mbox{m}$ for all modules produced (Fig.~\ref{CMS_gantry_precision}). 

At the start of the development of the production process the goal was set to achieve an accuracy better than $10~\micro\mbox{m}$ between sensor and frame, and $20~\micro\mbox{m}$ between the FE hybrid and the frame. Both was achieved. In order to do so, the gantry had to be elaborately calibrated before the production process, down to $2~\micro\mbox{m}$ precision in the $50\times50~\mbox{cm}^2$ $xy$-plane, using custom made plates with markers, allowing for an on-the-fly correction to the gantry's movement during the production.

\subsection{Assembly and metrology of back-to-back modules}
The sensors of all modules in the current and future strip tracker are highly processed and segmented on the front-side, while a uniform metalization layer is covering the complete back-side of the sensors. Since the sensors need to be mounted back-to-back, no optical common reference for both sensors is available when surveying the assembly only from the lower or upper side. This prohibits a similar construction scheme as used for the current modules.

\begin{figure}
\begin{center}
\includegraphics[width=.6\textwidth]{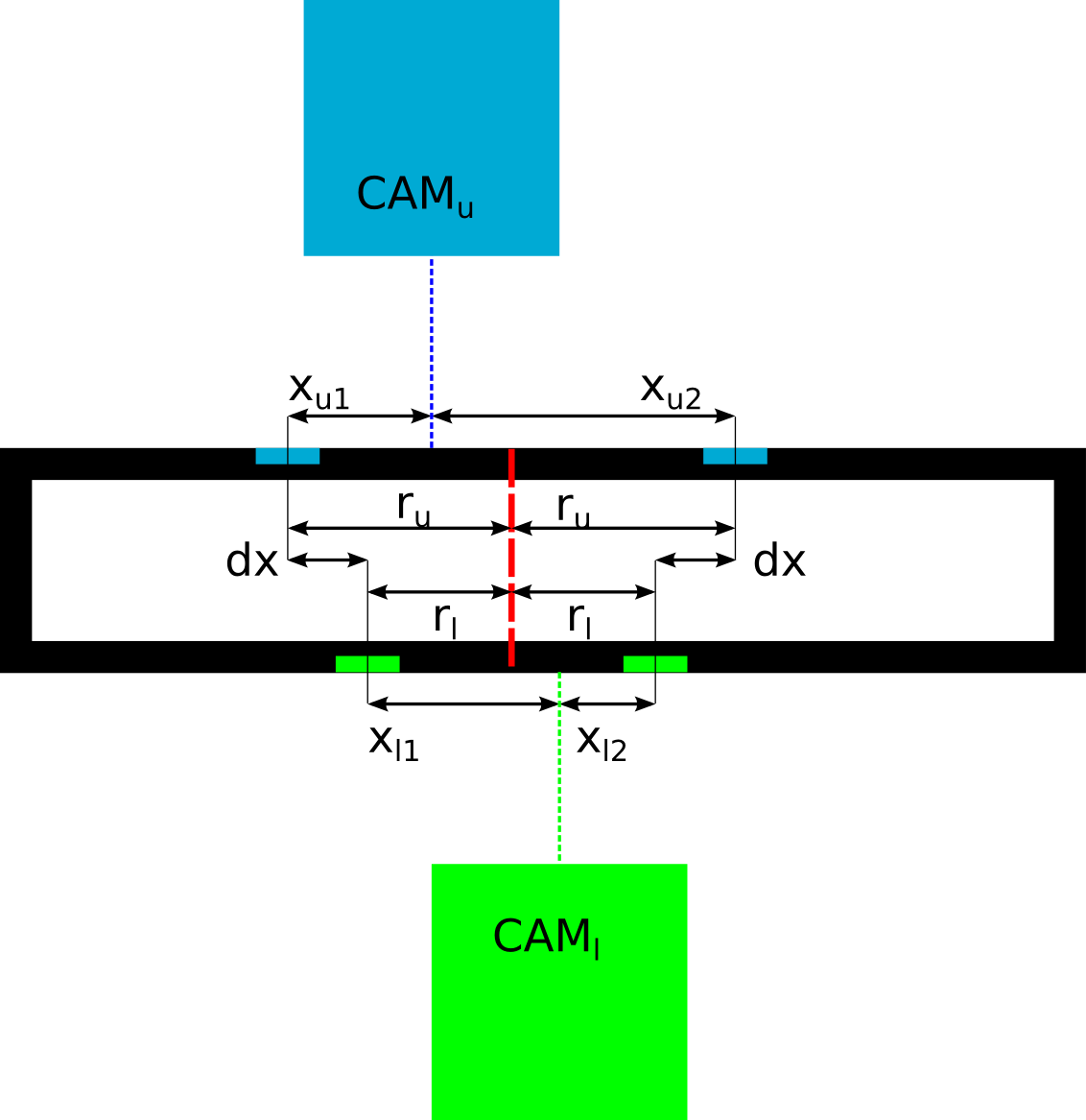}
\caption{Sketch of the double-sided measurement procedure, in one dimension. The rotation axis is shown in red. The blue and green spots on the top and bottom of the module (indicated by the black frame) symbolize alignment markers on the sensors. The cameras are displaced with respect to each other along the $x$ direction. In a first measurement the distances $x_{u1}$ and $x_{l1}$ between the markers and the camera axes are measured. After a rotation of the module by 180°, the distances $x_{u2}$ and $x_{l2}$ are obtained. The radial displacements $r_u$ and $r_l$ are given by the means of the $x$ values of the two measurements, respectively. The difference between the sensor marker positions is $dx = r_u - r_l$.}
\label{CMS_idea}
\end{center}
\end{figure}

The measurement of an opaque object from both sides requires either a complicated turning procedure, or a precise and stable alignment of two cameras installed above and below the object. However, misalignments of those cameras can be tolerated when the measurement is done twice, with a rotation of the module by 180° inbetween the two measurements. Taking the mean of those measurements, any (in-plane) misalignments cancel each other. This is illustrated in Fig.~\ref{CMS_idea}. Put differently, the axis of rotation acts as a common reference for both sides. The combination of measurements at all corners of the sensors allows for the determination of the displacement and rotation of the sensors with respect to each other.

\begin{figure}
\begin{center}
\includegraphics[width=.6\textwidth]{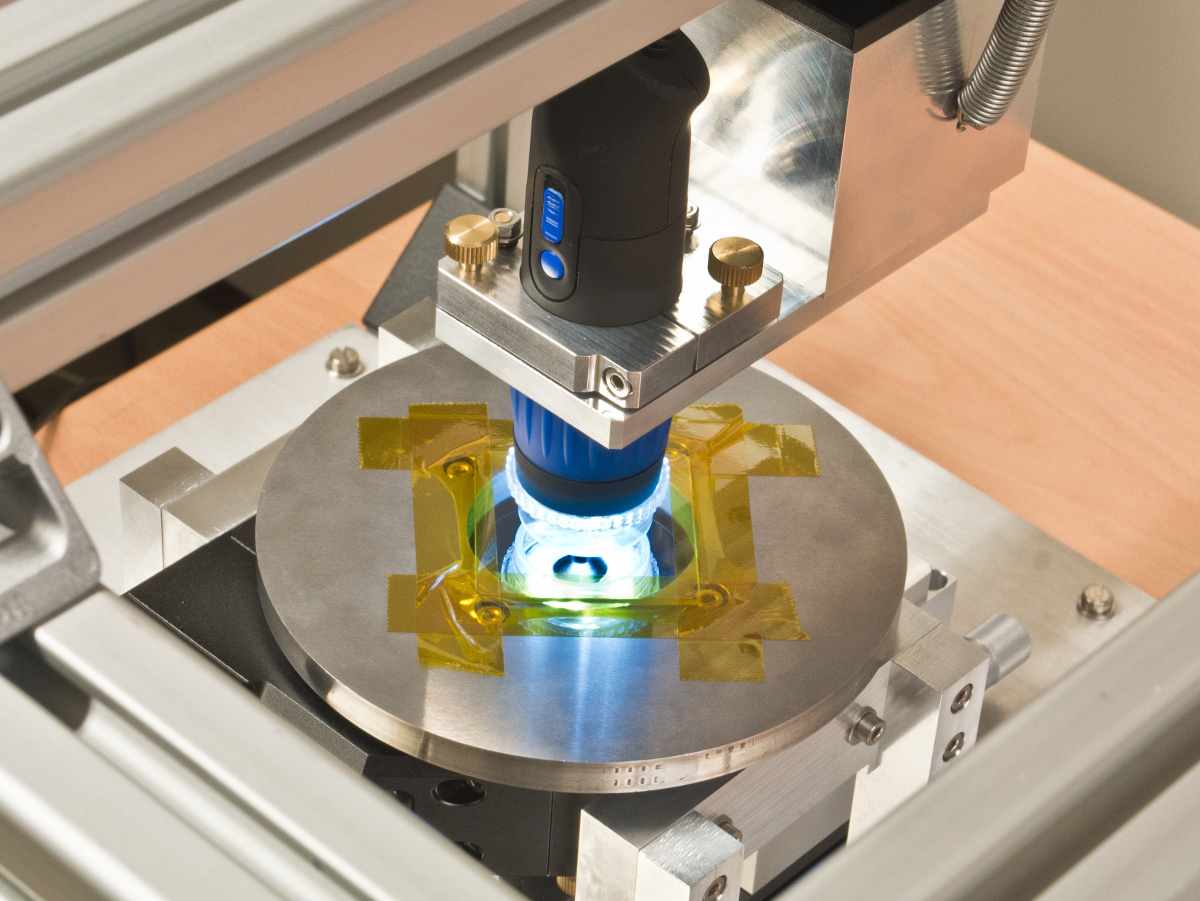}
\caption{Close-up view of the prototype double-sided metrology machine.}
\label{CMS_DSM}
\end{center}
\end{figure}

A small prototype machine using this concept was built as part of the work package 4 in order to demonstrate the feasibility of this concept at fairly low costs. A rotary stage provides the axis of rotation for the object under study, while two cameras, essentially facing each other, are surveying its upper and lower surface (Fig.~\ref{CMS_DSM}). 
A glass-plate with metal dots was used to demonstrate the precision. The same dot is measured from top and from below, so by construction the real displacement is zero. During these tests, a disagreement for radial measurements smaller than $2\,\micro\meter$ between the measurements at both surfaces was observed.

In order to allow for an automated complete measurement at several positions of a sensor assembly, the prototype was equipped with a lightweight but precise $xy$-stage, mounted on top of the rotary stage. Therefore the object under study can be moved with respect to the fixed rotation axis and camera assembly. First preliminary tests show a high stability of the rotation axis even under the additional load of the $xy$-stage. Because the instability of the rotation axis is the major source of uncertainties in this measuring scheme, we are optimistic that an adequate measurement accuracy will be achieved to qualify deviations in the $10\,\micro\meter$ range.

The metrology machine can be used during the module prototyping phase for qualification of the assembly precision of various assembly methods under study, and might serve as blue-print for module qualification during the final production run. It will be expanded to allow semi-automated assembly, providing measuring and positioning capabilities during the gluing of sensors and spacers. The addition of a glue-dispenser is planned as well.

\subsection{Assembly and metrology of double-sided petals}

The assembly of petals is conducted in several steps. First, hybrids and modules are built. Afterwards the petal cores are stuffed with modules.

The hybrid PCBs are produced at industry. The assembly of ASICs is performed afterwards at the institutes. This requires only a modest precision to serve proper wire-bonding and handling. The latter is done with pick-up tools touching centrally the ASIC surfaces. The same tools are used both for handling of fully tested and working hybrids and assembly of modules. This is exploited to cover partially the backside of the hybrids with glue and place them on the silicon-sensor surface. Also this step does not require high precision, but has to allow proper wire-bonding of strip pads to ASIC pads. Details on the assembly are also given in \cite{hauser}.
The hybrid assembly is planned to be by panels to allow short production times and mass production.
These assembly steps have been exercised in detail during the last three years.
In Freiburg and at DESY about 50 modules and 80 hybrids were built. Except one all are assembled correctly and electrical failures occured only on less than 10\,\%. Specific tooling allows for constant gluing heights when gluing ASICs onto hybrids. Variations of $\pm 15\,\micro\meter$ have been measured.
The hybrids show a good performance with a low mean noise of 375\,electrons and long-term noise stability over more than 80\,hours.
Glue height measurements, as well as optical and electrical tests, were continously performed during assembly. Modules show also a low mean noise, agreeing with expectations from calculations. Constant gluing heights ($100\pm 15\micro\meter$ for modules) were achieved. Thus, the prototyping with silicon sensors of 300\,$\micro\meter$ thickness is well controlled and shows very good results.
The procedure is also expected to work during mass production after proper training at production sites. 

Some institutes outside of Germany just started to investigate on hybrid assembly in industry. It might be possible. However, module assembly might be very difficult outside of institutes since detailed testing of components is required between assembly steps.
One possibility to speed up mass production is the use of glues with short curing times. However, the current prototyping is expected to be applicable to mass production. The investigations on glues with short curing times will be layed out below. The metrology of modules, which have a maximal size of 10\,cm\,$\times$\,10\,cm, is done using measurement microscopes.

\begin{figure}
\begin{center}
\includegraphics[width=.6\textwidth]{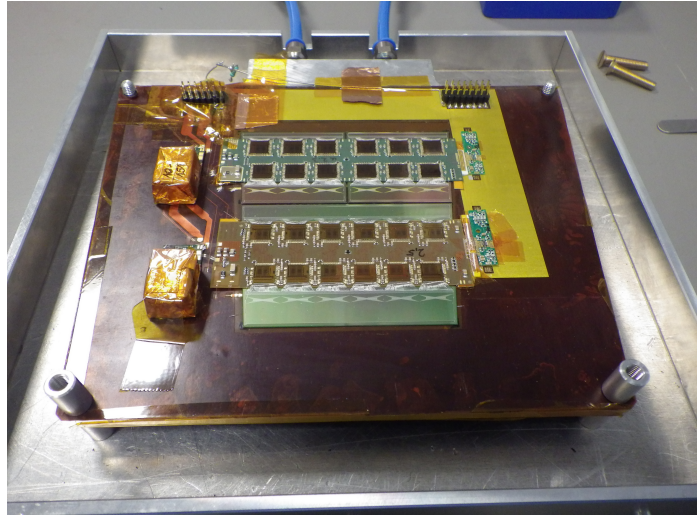}
\caption{Photo of a petalet seen from top. Two further modules are glued to the backside.}
\label{petalet1}
\end{center}
\end{figure}

A much higher precision is needed for mounting modules on cores. Essential is a system which allows referencing frontside to backside.
The prototyping for the end-cap region is ongoing with so-called petalets.
Petalets have a size of about 20$\times$17\,cm$^2$ and are made out of similar materials as petals.
They have three silicon sensors on each side. Depending on the layout of the readout, they have two or three hybrids per side. A photo of a petalet is shown in Fig.\,\ref{petalet1}. It shows two modules glued on a carbon core. Power is provided using DC-DC converters and readout is performed with specific boards.

\begin{figure}
\begin{center}
\includegraphics[width=.9\textwidth]{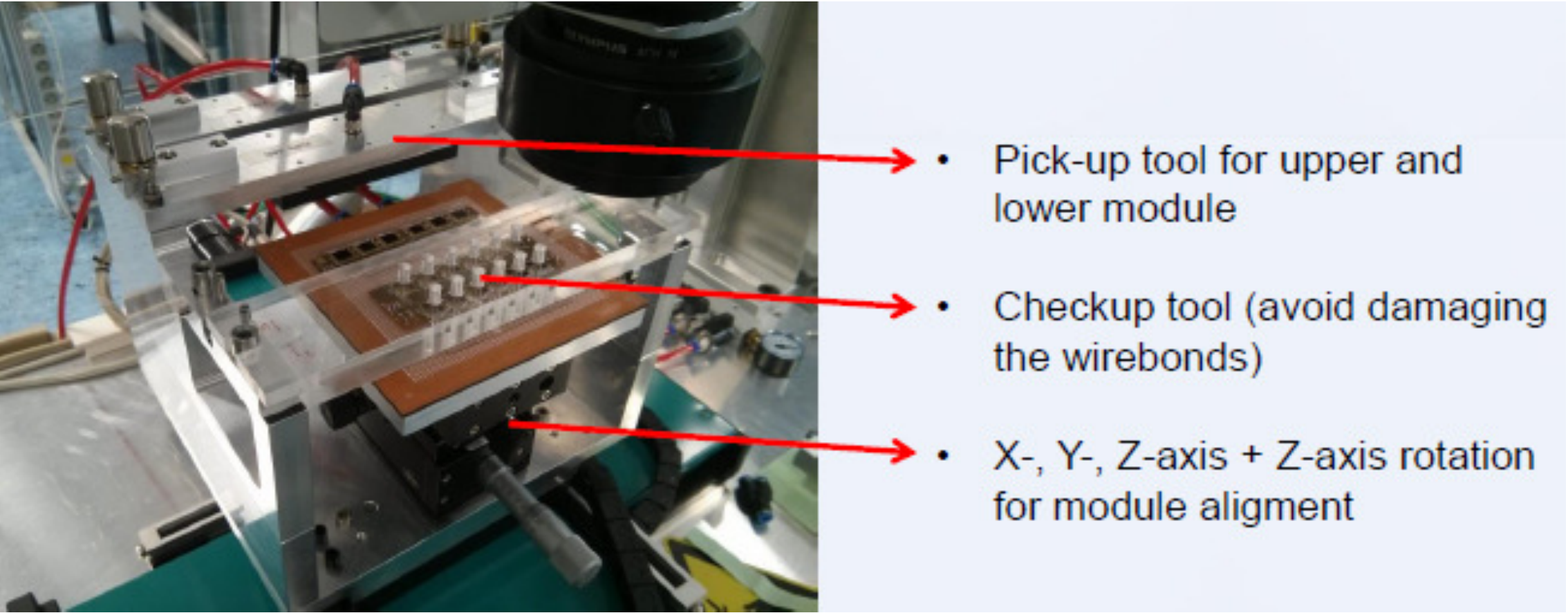}
\caption{Assembly tool for mounting modules on petalet cores at DESY.}
\label{desystage}
\end{center}
\end{figure}

Currently, the development of the two-sided assembly of petalets and petals is still ongoing.
First tests were performed by gluing manually. Only very rough precision was achieved.
A set-up using a microscope and positioning stages is being developed at DESY (Fig.~\ref{desystage}). It shows the pick-up tool to lift modules, $x$-, $y$-, and $z$-stages, a rotation stage and the module placement area. 
A perspex plate with specific holes (\textquotedblleft checkup tool\textquotedblright in the figure) is used for optical inspection before lowering the pick-up tool (module), avoiding damage to any wire-bonds.

A first petalet was successfully assembled using the tool. In a second step the module positions have to be surveyed which is intended to be done using a measurement microscope. The referencing might be done using a frame which fixes the petalet position precisely. It will have reference markers on top and bottom which are visible from both sides.  

Although the institutes involved in the work package 4 are mainly working on the upgrade of the end-caps, it should be mentioned that for the barrel region a motorized $x$-$y$-$z$ stage assembly machine was developed at the Rutherford Appleton Laboratory, Didcot, UK. It exploits laser alignment systems and software driven position optimization.
Moreover, the staves are fixed in precisely machined aluminium stages which have fiducials visible from top and bottom. The system is designed to achieve a precision of $\pm 5\,\micro\meter$  perpendicular to
the strips and  $\pm 50\,\micro\meter$ along the strips.
A similar system will be needed for the end-caps and work is ongoing to develop a machine with industrial partners.
Within the work package 4, possibilities were evaluated to buy such an assembly machine at industry. Several companies, namely H\"acker Automation, Schwarzhausen (www.haecker-automation.com), Baumann GmbH (www.baumann-automation.com), Ficontec (www.fi\-contec.com), XENON Automatisierungstechnik (www.xenon-automation.com), SUESS MicroTec (www\-.suss.com) and Finetech (www.finetech.de) were investigated. The companies H\"acker and Ficontec were contacted and possible assembly machine solutions have been discussed and are followed up with the idea of developing a small-scale prototype machine.

%Assembly: Methods/tools for alignement during assembly, use of fiducials etc., consecutive steps of assembly, availability and reliability for mass/automated production (asembly speed)
% Metrology: measurements tools, tools for determination of precision of two-sided objects and of large objects (~ 1m length of Petal)
% Change in precision due to environments (temperature, humidity), resulting necessity of stability, stiffness
% Survey on possible rework procedures ?
%Logistics in large scale production ?

\begin{figure}
\begin{center}
\includegraphics[width=.7\textwidth]{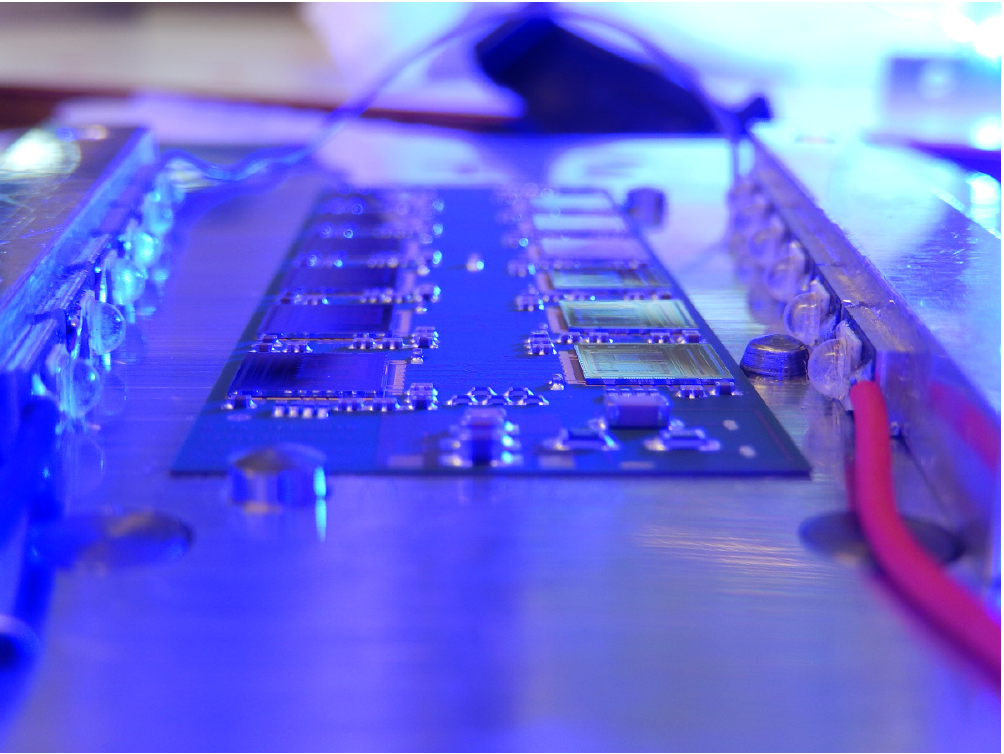}
\caption{Hybrid during curing of the glue, used to glue the ASIC to the hybrid, with UV light.}
\label{uvglue}
\end{center}
\end{figure}

Another study conducted within the work package 4 is the above mentioned investigation of UV cure glues with very short curing time.
They would allow a faster production without requiring parallel production chains at individual institutes, and reduce the costs for adhesives in module production by 95\,\%. 
The use of commercially available glues (six UV cure glues by three different producers and one glue pad) to connect the readout chips to the hybrid was studied in detail. The construction of prototypes with UV cure glues/glue pads and a subsequent series of tests (thermal conductivity, thermal cycling, irradiation, shear strength) resulted in three good UV cure candidates: DYMAX 3013, DYMAX 6-621 and LOCTITE 3525. First circuit board prototypes have been produced using UV cure glues and tests have shown good electrical performances. One circuit board prototype was already glued to a sensor.
A hybrid during curing with UV light can be seen in Fig.\,\ref{uvglue}.

In parallel, a second study was started to investigate the use of the three good UV cure glues between the FE hybrid and the sensor. Silicon strip sensor prototype miniatures were connected to circuit board pieces using either a UV cure glue (DYMAX 3013, DYMAX 6-621 and LOCTITE 3525) or the epoxy glue currently in use (Epolite FH 5313). The effects of every glue on the sensor were investigated by comparing sensor characteristics (leakage current, breakthrough voltage, interstrip capacity and resistance) before and after gluing. All sensors will be subjected to further tests (thermal cycling, irradiation), after which their characteristics will be compared again. Options to use UV cure glues for module mass production are currently under investigation. A first programme for the use of UV cure glues in an automatic glue dispenser has been successfully tested at Birmingham and UV-LED tools for curing have been developed and successfully tested at HU Berlin/DESY.

\subsection{Conclusions on automated precision assembly procedures}

The current silicon strip trackers of the CMS and ATLAS experiments were built with high precision and are performing very well. Knowledge of their assembly was collected in this work package, aiming for transfer of knowlegde for the assembly and survey of the upgrade strip trackers.
The plans for building modules and larger structures for the upgrade strip trackers of both experiments were compared. Currently, prototyping is ongoing and both module assembly and the assembly of larger structures and their survey is being investigated. Collaborators from the CMS experiment built a prototype for metrology of double-sided modules within work package 4. Here sensors need to be mounted back-to-back precisely, and the prototype setup showed its ability to reach a measurement precision below 10\,$\micro\meter$. Collaborators from the ATLAS experiment built successfully first circuit boards (hybrids) and modules with the required precision. Those need to be mounted precisely onto carbon structures. This procedure was also exercised using tools with optical and manual alignment. According to the experience from gantry assembly in the CMS experiment, it seems to be possible to assemble modules onto staves and petals using gantries. This is being further investigated. A further study towards mass production was covering UV curing glues to reduce the curing time of hybrid production. Several suitable glues were found.
Moreover assembly possibilities in industry were evaluated.
The work package lead to a fruitful exchange of knowledge between the collaborators of the ATLAS and CMS experiments, and the survey of a double-sided petalet for the ATLAS experiment using the CMS survey tool is planned. 

\newpage

\section{Irradiations}
This work package provided coordination and funds for irradiation projects within the framework of R\&D for future silicon trackers. Members of this Helmholtz Alliance project were allowed to apply for funds covering beam time at the Karlsruhe Cyclotron or travel expenses to other radiation sources, for which no direct access costs are claimed.
The applications have been checked for relevance and feasibility. The individual irradiation projects were initially limited to four hours of beam time to allow several applications by many groups. 

During the project lifetime nine applications were received. All projects have been accepted and executed. Table~\ref{irradiations} 
gives a summary of the irradiation projects. The individual projects are shortly described below:

\begin{table}
	\centering
\begin{tabular}{llcccc}
\hline
\textbf{Title} & \textbf{Time / h} & \textbf{Dosimetry }  \\
\hline
Materials for DC-DC Powering, CMS & 2.60 & 4        \\
ATLAS glue tests I & 3.25 & 2      \\
Transistors for ATLAS Tracker & 2.97 & 2      \\
Irradiation of HPK mini sensors for ATLAS end-cap upgrade & 1.17 & 1      \\
Thermally conductive glues for ATLAS & 0.83 & 1       \\
p-stop isolation study for CMS Tracker & 1.08 & 1       \\
Irradiation of CMS 2S prototype sensor &  0.42 & 1      \\
ATLAS glue tests II & 3.98 & 2    \\
\hline
\textbf{Sum} & \textbf{16.30} & \textbf{14}  \\
\hline
\end{tabular}
\caption{\label{irradiations}Summary of irradiation projects conducted at ZAG, KIT. The column labelled "Dosimetry" refers to the number of dosimetry measurements that were performed.}
\end{table}

\begin{description}
	\item[Materials for DC-DC Powering for CMS] All components of DC-DC converter prototypes for the CMS Tracker upgrade have been validated at a fluence of $4\times 10^{14}\;\mathrm{n}_{\mathrm{eq}}\mathrm{cm}^{-2}$. Among those are passive electronic components, fuses, electro-magnetic shields and whole assemblies. The samples were irradiated in six sessions to a total of 2.60 hours.
	\item[ATLAS glue tests I] Within the ATLAS Tracker Upgrade R\&D alternative glues to silver epoxy are being evaluated (thermal conductivity, mechanical stability, glue strength, ...). This requires radiation hardness up to a fluence of $2\times 10^{15}\;\mathrm{n}_{\mathrm{eq}}\mathrm{cm}^{-2}$. The samples consist of pure glue drops and chips glued to hybrids. They were irradiated in two sessions to a total of 3.25 hours.\\
	%Report given at 7th Detector Workshop of the HHA, March 2014 (\url{https://indico.desy.de/getFile.py/access?contribId=26&sessionId=2&resId=0&materialId=slides&confId=9389}).
	\item[Transistors for the ATLAS Tracker] The high-voltage (HV) biasing concept of the ATLAS Strip Tracker should minimize the number of cables. One option would be switching of HV using transistors. Several transistors have been irradiated to fluences up to $1\times 10^{15}\;\mathrm{n}_{\mathrm{eq}}\mathrm{cm}^{-2}$ under bias. The preparations of these irradiations were quite time consuming due to the fixation of about 20 bias lines and the need to irradiate from both sides due to the thickness of the devices. The samples were irradiated in four sessions for a total of 2.97~hours.
	\item[Irradiation of HPK mini sensors for the ATLAS end-cap upgrade] p-type silicon strip sensors of 320~$\micro\meter$ thickness have been irradiated to $5\times 10^{14}\;\mathrm{n}_{\mathrm{eq}}\mathrm{cm}^{-2}$, $1\times 10^{15}\;\mathrm{n}_{\mathrm{eq}}\mathrm{cm}^{-2}$ and $2\times 10^{15}\;\mathrm{n}_{\mathrm{eq}}\mathrm{cm}^{-2}$. The samples were irradiated in one session to a total of 1.17~hours.
	\item[Thermally conductive glues for ATLAS] This project aims at comparing SE4445 (silicone gel) and SE4468 (polydimethyl-siloxane adhesive) from Dow Corning as glue between silicon sensor and Kapton foil. The irradiation fluence was $2\times 10^{15}\;\mathrm{n}_{\mathrm{eq}}\mathrm{cm}^{-2}$. The samples were irradiated in one session for a total of 50~minutes.
	\item[p-stop isolation study for the CMS Tracker] p-type silicon strip sensors require measures to isolate the strips against each others. One way is the implantation of a highly doped p-stop ring around each strip (atoll design). This study investigates the impact of the p-stop doping concentration before and after irradiation up to $1.5\times 10^{15}\;\mathrm{n}_{\mathrm{eq}}\mathrm{cm}^{-2}$. The samples were irradiated in one session for a total of 1.08~hours.
	\item[Irradiation of CMS 2S prototype sensor] The CMS Strip Tracker will make use of correlated hits in modules with stacked sensors to discriminate for the particle momentum (e.g. large displacement for low momentum tracks in the strong magnetic field). A sensor prototype with close to final layout parameters has been irradiated to $7\times 10^{14}\;\mathrm{n}_{\mathrm{eq}}\mathrm{cm}^{-2}$ to study the performance attached to the current prototype of the foreseen binary readout chip. The sample was irradiated in one session for a total of 25~minutes.
	\item[ATLAS glue tests II] This second project systematically investigates the use of UV curing glues as alternatives to epoxy based ones for the connections readout chip to PCB and PCB to sensor. Irradiations were performed at Karlsruhe (23~MeV protons) for 3.98 hours.
	\item[Irradiations of surface structures with 800~MeV protons at Los Alamos] The aim of this project is to investigate the electrical field at the interface with irradiated structures that allow to determine the surface damage (MOS, gate-controlled diodes GCD). In combination with sensors that allow the measurement of the damage in the silicon crystal (diode, strip sensors), the generation of damage and the interplay between surface and bulk damage is studied. Irradiation at Los Alamos is free of charge, but the travel expenses for two persons for local preparations and shifts have been partially remunerated.
\end{description}

\section{Conclusions}
The PETTL project was very successful in establishing a fruitful exchange between the German groups active in the upgrades of the ATLAS and CMS tracking systems, and in strengthening the links between DESY and the University groups. The exchange of the experience gained and lessons learned during the construction of the current ATLAS and CMS tracking detectors has been extremely useful. It will help to improve the quality of the future tracking detectors and to make their construction more efficient. Lots of data on the radiation hardness of silicon sensors has been acquired by both collaborations. The joint analysis of this data within the PETTL project has considerably improved the mutual comprehension of the data sets and has led to a better understanding of radiation effects and their simulation. An important goal for the new tracking detectors is low mass design which has been reviewed and compared between ATLAS and CMS. Finally, the requirements and the solutions proposed for the precision assembly of the ATLAS and CMS tracking detector modules have been analyzed in the light of the experience with module assembly for the current trackers. Ideas for double sided metrology and novel gluing concepts have been explored. Three workshops have been organized within the PETTL project. They were essential to start common activities, to build a network between the participating groups, and to present and discuss results. The collaboration established through PETTL will continue beyond the formal end of the project.

\end{document}